\documentclass[prd,aps,twocolumn,floats,floatfix,nofootinbib]{revtex4-1}
\usepackage{graphicx}
\usepackage{amsmath,amssymb}
\usepackage{physics}
\usepackage{subfigure}
\usepackage{appendix}
\usepackage[table]{xcolor}
\usepackage{hyperref}
\usepackage{booktabs} % For \toprule, \midrule and \bottomrule
\usepackage{longtable} 
\usepackage{acronym}
\usepackage{enumerate}
\usepackage{cellspace}

\definecolor{mangotango}{rgb}{1.0, 0.51, 0.26}

\newcommand{\comment}[1]{}

\def \qupm {\textsc{qu-pm}}
\def \nopm {\textsc{no-pm}}
\def \free {\textsc{free-pm}}
\def \stwo {$\mathrm{Source2}_{\mathrm{[qu-pm]}}$}
\def \sth {$\mathrm{Source3}_{\mathrm{[qu-pm]}}$}
\def \sone {$\mathrm{Source1}_{\mathrm{[qu-pm]}}$}

\newacro{GW}{gravitational-waves}
\newacro{BNS}{binary-neutron-star}
\newacro{ET}{Einstein Telescope}
\newacro{CE}{Cosmic Explorer}
\newacro{PSD}{power spectral density}
\newacro{SNR}{signal-to-noise-ratio}
\newacro{PE}{parameter estimation}
\newacro{BBH}{binary-black-hole}
\newacro{MCMC}{Markov chain Monte Carlo}
\newacro{BH}{black hole}

\begin{document}

\title{Unraveling information about supranuclear-dense matter from the 
complete binary neutron star coalescence process using future gravitational-wave detector networks}

\author{
Anna Puecher$^{1,2}$,
Tim Dietrich$^{3,4}$, 
Ka Wa Tsang$^{1,2,5}$,
Chinmay Kalaghatgi$^{1,2,6}$,
Soumen Roy$^{1,2}$,
Yoshinta Setyawati$^{1,2}$,
Chris Van Den Broeck$^{1,2}$
}

\affiliation{$^1$Nikhef -- National Institute for Subatomic Physics, Science Park 105, 
1098 XG Amsterdam, The Netherlands \\
$^2$Institute for Gravitational and Subatomic Physics (GRASP), Utrecht University, 
Princetonplein 1, 3584 CC Utrecht, The Netherlands  \\
$^3$Institut f\"{u}r Physik und Astronomie, Universit\"{a}t Potsdam,
Haus 28, Karl-Liebknecht-Str.~24/25, 14476, Potsdam, Germany \\
$^4$Max Planck Institute for Gravitational Physics (Albert Einstein Institute), Am M\"{u}hlenberg 1, Potsdam, Germany \\
$^5$Van Swinderen Institute for Particle Physics and Gravity, University of Groningen, 
Nijenborgh 4, 9747 AG Groningen, The Netherlands\\
$^6$Institute for High-Energy Physics, University of Amsterdam, Science Park 904, 1098 XH Amsterdam, 
The Netherlands}

\date{\today}

%===================================================
% Abstract

\begin{abstract}
Gravitational waves provide us with an extraordinary tool to study the matter inside neutron stars. 
In particular, the postmerger signal probes an extreme temperature and density regime and will help 
reveal information about the equation of state of supranuclear-dense matter. 
Although current detectors are most sensitive to the signal emitted by binary neutron stars 
before the merger, the upgrades of existing detectors and the construction of the next generation 
of detectors will make postmerger detections feasible. For this purpose, we present a new analytical, 
frequency-domain model for the inspiral-merger-postmerger signal emitted by binary neutron stars 
systems. The inspiral and merger part of the signals are modeled with \texttt{IMRPhenomD$\_$NRTidalv2}, 
and we describe the main emission peak of postmerger with a three-parameter Lorentzian, using two 
different approaches: one in which the Lorentzian parameters are kept free, and one in which we model 
them via quasi-universal relations. We test the performance of our new complete waveform model in 
parameter estimation analyses, studying simulated signals obtained from both our developed model and by injecting 
numerical relativity waveforms. 
We investigate the performance of different detector networks to determine the improvement that future 
detectors will bring to our analysis. We consider Advanced LIGO+ and Advanced Virgo+, KAGRA, and 
LIGO-India. We also study the possible impact of a detector with high sensitivity in the 
kilohertz band like NEMO, and finally we compare these results to the ones we obtain with 
third-generation detectors, the Einstein Telescope and the Cosmic Explorer.

%We compare the results for the LIGO-Virgo network with aLIGO+ sensitivity to a network including also LIGO-India and KAGRA. We also study the possible improvement given by the high-frequency detector NEMO, and we finally compare results to what we will obtain for the same sources with third-generation detectors.
\end{abstract}

\maketitle

%===================================================
\section{Introduction}
\label{sec:introduction}

Neutron stars (NSs) can reach extremely high densities, creating conditions that cannot 
be reproduced by laboratory experiments. Hence, they provide a perfect environment to study 
supranuclear-dense matter and its Equation of State (EoS). Until a few years ago, 
the study of NSs was limited to electromagnetic (EM) observations, but since the 
first detection of a gravitational wave (GW) signal from a binary neutron star (BNS), 
GW170817 \cite{LIGOScientific:2017vwq}, GWs provide new ways to study NSs and their mergers. 
Since the EoS determines the NS's macroscopic properties, such as its mass, radius, and tidal 
deformability, it can be constrained by measuring the imprint it leaves in the GW signal 
emitted during the coalescence \cite{Dietrich:2020eud,Chatziioannou:2020pqz}. 

Up to now, Advanced LIGO~\cite{LIGOScientific:2014pky} and Advanced Virgo~\cite{VIRGO:2014yos} detected two BNS systems, GW170817 \cite{LIGOScientific:2017vwq, LIGOScientific:2018hze} 
and GW190425 \cite{LIGOScientific:2020aai}. These detections already allowed 
to put constraints on the supranuclear-dense matter EoS, which was possible since the GW signal 
emitted during the inspiral phase provides information about the EoS through tidal deformability 
measurements~\cite{Dietrich:2020eud,Hinderer:2009ca, Damour:2012yf, DelPozzo:2013ala,Lackey:2014fwa, 
Agathos:2015uaa, Dietrich:2018uni, Kawaguchi:2018gvj, Hinderer:2007mb, Damour:2009wj}. 
While the uncertainty on current measurements is still large, the higher sensitivies of 
future generation detectors such as the Einstein Telescope (ET) 
\cite{Punturo:2010zz, Maggiore:2019uih, Freise:2008dk, Hild:2009ns,Sathyaprakash:2011bh,Pacilio:2021jmq,Gupta:2022qgg} 
or the Cosmic Explorer (CE) \cite{Reitze:2019iox, Evans:2021gyd} will significantly improve them.

In addition to a more detailed analysis of the inspiral, 3rd generation (3G) GW detectors such 
as ET and CE are also expected 
to detect GWs from the postmerger phase of the BNS coalescence~\cite{Koppel:2019pys,Bauswein:2013jpa,Baiotti:2008ra,Bernuzzi:2020tgt,Baiotti:2016qnr}. 
This is of special interest, since the postmerger probes an even higher different density and temperature regime than the inspiral. 
While during the inspiral only densities up to the central density of the individual stars are probed, which corresponds to about 3 to 4 times nuclear saturation density, 
the postmerger phase probes densities even beyond five times nuclear saturation density, cf.~Fig.~1 of~\cite{Pang:2022rzc}. 
In addition, also temperatures of about 50 MeV are reached during the postmerger phase, which is large enough so that the effect of different transport coefficients will start to impact the data \cite{Hammond:2021vtv,Raithel:2021hye,Most:2022yhe}.

Unfortunately, postmerger studies pose numerous challenges. Firstly, 
the amplitude of the GW strain of the postmerger part of the 
observed GW signal is expected to be weaker than the inspiral 
one~\cite{Kastaun:2014fna, Bauswein:2012ya, Takami:2014tva, Bernuzzi:2015rla, Bauswein:2015yca}. 
Secondly, at higher frequencies the detectors' sensitivity drops due to 
quantum shot noise.
For these reasons, it is not surprising that the dedicated searches for GWs 
emitted by a possible remnant of GW170817~\cite{LIGOScientific:2017fdd,LIGOScientific:2018urg} 
found no evidence of such a signal, and showed that with the sensitivity of Advanced LIGO and 
Advanced Virgo the source distance should have been at least one order of magnitude less for the postmerger signal to be detectable.
Finally, postmerger physics includes thermal effects, magnetohydrodynamical instabilities, 
neutrino emission, dissipative processes, and possible phase transitions 
\cite{Bauswein:2018bma, Most:2018eaw, Siegel:2013nrw, Alford:2017rxf, Radice:2017zta, Shibata:2017xht, DePietri:2018tpx}, 
which make the postmerger particularly difficult to model, but, on the other hand, 
allow us to investigate a variety of interesting physical processes.
Because of the complexity of the evolution, the study of the postmerger relies heavily on 
numerical-relativity (NR) simulations, which, however, are also limited due to their high 
computational cost and the fact that it is currently not possible to take into account all 
the physical processes that influence the postmerger. 

Nonetheless, previous studies based on NR simulations showed some common key features of the postmerger GW spectrum, finding in some cases universal relations with the NS properties \cite{Bauswein:2011tp, Bose:2017jvk, Clark:2014wua, Takami:2014zpa, Rezzolla:2016nxn, Bernuzzi:2015rla, Bauswein:2012ya, Hotokezaka:2013iia, Bauswein:2014qla, Takami:2014tva, Bauswein:2015yca, Chatziioannou:2017ixj, Lioutas:2021jbl}, and some efforts have been made also to construct full inspiral, merger and postmerger models for BNS coalescences.
Also morphology-independent analyses of the postmerger GW signal have been proposed in \cite{Clark:2014wua, Clark:2015zxa, Chatziioannou:2017ixj}, while in \cite{Easter:2018pqy} a hierarchical model to generate postmerger spectra was developed.
With a different approach, \cite{Breschi:2019srl,Easter:2020ifj,Soultanis:2021oia} construct analytical models for the postmerger signal, based on features found in NR simulated waveforms. Breschi \textit{et al.} in \cite{Breschi:2022xnc} proposed a frequency-domain model for the postmerger, built with a combination of complex Gaussian wavelets, and showed in \cite{Breschi:2022ens} how this model performs using a 3G detector network. Wijngaarden \textit{et al.} \cite{Wijngaarden:2022sah} build a hybrid model, using analytical templates for the premerger phase and a morphology-independent analysis, based on sine-Gaussian wavelets, for the postmerger one. %This approach allows to study the BNS signal with all the data available, and also to perform consistency tests between the pre- and postmerger results.

Following similar ideas, in this paper we construct a phenomenological frequency domain model for 
the entire BNS coalescence consisting of the inspiral, merger, and postmerger phase. Our 
final aim is it to employ the developed model for parameter estimation analyses. To model 
the coalescence during the inspiral up to the merger, we rely on \texttt{IMRPhenomD$\_$NRTidalv2} 
\cite{Dietrich:2019kaq}. The postmerger phase is modelled with a three-parameter Lorentzian 
describing the main emission peak of its spectrum, following Tsang \textit{et al.\ } \cite{Tsang:2019esi}. 
For the Lorentzian, we use two different approaches: in one case, we compute the parameters 
from quasi-universal relations, describing them as a function of the BNS's properties, in 
the other one, we treat them as free parameters. 
Both versions can be directly employed by existing parameter estimation pipelines; see e.g.~\cite{Ashton:2018jfp,Romero-Shaw:2020owr}.

%The advantage of having a full analytical model is that it can be directly employed by existing parameter estimation pipelines; see e.g.~\cite{Ashton:2018jfp,Romero-Shaw:2020owr}. 

%\textcolor{red}{There are also some more recent works of Andreas Bauswein that should be added.}

This paper is structured as follows. In Sec.\ref{sec:methods} we describe how our model is built, the 
methods used for parameter estimation, and the detectors we consider. Results are shown in 
Sec.~\ref{sec:results}, and conclusions are presented in Sec.~\ref{sec:conclusions}. 
Appendix~\ref{sec:free_parameter} shows the results obtained specifically with our postmerger model with free Lorentzian parameters, in Appendix~\ref{sec:pe} we discuss general parameter estimation with future detectors, and in Appendix~\ref{sec:relbin_app} we provide more details about the validity and settings of the methods employed for our study.

%Appendix~\ref{sec:mismatch} is devoted to mismatch calculations between our models and the 
%NR waveforms, while in Appendix~\ref{sec:free_parameter} we show the results 
%obtained specifically with our postmerger model with free Lorentzian parameters.

\section{Methods and Setup}
\label{sec:methods}

We construct a frequency-domain waveform model to describe the full inspiral, merger, and postmerger of a BNS coalescence. In this section, we describe how we model the postmerger part of the signal, and how we connect it to the inspiral-merger model to obtain the full waveform. We then describe the framework used for data analysis, explaining how we speed up parameter estimation using relative binning, the analysis setup, the BNS sources that we study, and the employed detector networks to determine to what extent future detector networks will enable postmerger studies. 

\subsection{Inspiral-merger-postmerger model construction}
\label{sec:construction}

Multiple studies have shown that the postmerger GW spectrum includes various strong 
peaks~\cite{Rezzolla:2016nxn, Bauswein:2011tp, Stergioulas:2011gd, Bauswein:2012ya, Hotokezaka:2013iia, Bauswein:2014qla, Takami:2014zpa, Bauswein:2015yca, Takami:2014tva, Bernuzzi:2015rla}. 
For simplicity, we limit ourselves to the main emission peak at a frequency $f_2$, 
which corresponds to the dominant GW frequency; see e.g.~\cite{Bauswein:2011tp}. 
Following this approach, the postmerger can be described in time domain by a simple damped 
sinusoidal waveform~\cite{Tsang:2019esi}, whose Fourier transform is a Lorentzian. 
Therefore, in frequency domain, we model the postmerger with a three-parameter Lorentzian
\begin{equation}
h_{22}(f) = \frac{c_0 c_2}{\sqrt{(f-c_1)^2 + c_2^2}} e^{-i \arctan{\left(\frac{f-c_1}{c_2}\right)}}, 
\label{eq:lorentzian}
\end{equation}
where $c_0$ corresponds to the maximum value, $c_1$ to the dominant emission frequency $f_2$, 
and $c_2$ to the inverse of the damping time, which sets the Lorentzian's width.

We determine the coefficients $c_i$ with two different approaches: 
(I) we treat them as free parameters, and try to measure $c_0, c_1$, and $c_2$ together with the 
other BNS's properties; and (II) we compute the $c_i$ coefficients from quasi-universal relations 
that describe them as functions of the system's parameters. 
Depending on its properties and EoS, a given BNS could undergo a prompt collapse to a black hole (BH), 
hence without a postmerger emission. In this scenario, while in case (I) we expect that the values 
recovered for the free parameters reflect the absence of a postmerger signal, in (II) the 
quasi-universal relations employed might lead to a bias in the estimation of the binary's 
intrinsic parameters. For this reason, we ideally want to use the Lorentzian model with 
quasi-universal relations only when we know that a postmerger emission is present. 
Since the threshold mass for a prompt collapse is EoS dependent and still unknown, 
following~\cite{Breschi:2019srl} we assume that a BNS system undergoes prompt collapse if the 
tidal polarizability parameter $\kappa_{2}^{T}$ is lower than a threshold value 
$\kappa_{\mathrm{thr}} = 40$. The quantity $\kappa_{2}^{T}$ is defined as
\begin{equation}~\label{eq:kappa2t_thrs}
\kappa_{2}^T = {3} \left[  \Lambda_{2}^{A} (X_{A})^{4} X_{B} + \Lambda_{2}^{B} (X_{B})^{4} X_{A} \right],
\end{equation}
where $\Lambda_2^j = \frac{2}{3} k_2 \left( R_j/M_j\right)^5$ with $j\in \{A,B\}$ are the dimensionless 
tidal deformabilities, and $X_j = M_j / M$. Here $k_2$ is the dimensionless $\ell = 2$ Love number, 
$R_j$ and $M_j$ are respectively the radius and gravitational mass of the individual stars, 
and $M = M_A + M_B$ is the BNS's total mass\footnote{See also~\cite{Kolsch:2021lub} 
for more updated relations which were not yet available when we started our work.}. 

%\td{Please check the following, I think Breschi has a typo in his paper.}
%Note that this definition of $\kappa_{2T}$ differs from the one used in \texttt{IMRPhenomD$\_$NRTidalv2}\cite{Dietrich:2019kaq} by a factor 2. \td{Is this true? Why do we d this then?}

\subsubsection{Quasi-universal relations for the Lorentzian parameters}
\label{sec:parameter_models}

For the approach introduced as method (II), we use quasi-universal relations, 
i.e.~phenomenological relations that are independent of the EoS, to constrain the 
coefficients $c_i$ in Eq.~\eqref{eq:lorentzian}. This provides a direct connection 
between the Lorentzian coefficients and the BNS's properties. 

Since the postmerger Lorentzian model extends the waveform used for inspiral and merger beyond its 
merger frequency $f_{\mathrm{merg}}$, a straightforward way to find the value of $c_{0}$ is 
by rescaling the amplitude of the \texttt{IMRPhenomD$\_$NRTidalv2} waveform at merger 
$\mathcal{A}_{\rm{NRTidalv2}}(f_\mathrm{merg})$ . Specifically, we use
\begin{equation}
c_0 = \sigma \times \mathcal{A}_0 \times \mathcal{A}_{\rm{NRTidalv2}}(f_{\rm{merg}}),
\end{equation}
where $\mathcal{A}_0$ is the mass and distance scaling factor employed in \texttt{IMRPhenomD}~\cite{Khan:2015jqa}. %\td{I don't fully understand the $\sigma$ discussion:} 
The prefactor $\sigma$ is added to obtain a better calibration to the NR waveforms, and we set $\sigma  = 10.0$, which gives the lowest mismatch values (the definition of mismatch and details about its computation are provided in Sec.~\ref{sec:mismatch}).

%The prefactor $\sigma$ is added in order to adjust the amplitude for the limited range of values that $c_2$ can take. As given in Eq.~\eqref{eq:lorentzian}, the overall Lorentzian amplitude depends also on $c_2$, and very low (high) values of $c_2$ cause the amplitude to be excessively spiked (wide) around the dominant postmerger frequency $f_2$. We set $\sigma  = 10.0$ because it gives the highest matches with NR waveforms (the definition of match and details about its computation are provided in Appendix ???\td{please add}).

%We choose to scale the amplitude by a given $\alpha$ to restrict the range of values that $c_{2}$ can take. For eg: a very low (high) value of $c_{2}$ could cause the amplitude to be highly spiked (wide) around $f_{2}$, same with the overall scale for the phase.  

Since $c_1$ represents the dominant postmerger oscillation frequency $f_2$, we resort to the fit 
in Eq.~(8) of \cite{Tsang:2019esi} 
\begin{equation}
M c_1(\zeta) = \beta \frac{1+A \zeta}{1 + B \zeta} ,
\label{c1_fit}
\end{equation}
with $\beta = 3.4285 \times 10^{-2}$, $A = 2.0796 \times 10^{-3}$, and $B = 3.9588 \times 10^{-3}$. 
The parameter $\zeta$ is
\begin{equation}
\zeta = \kappa_{\mathrm{eff}}^{T} -131.7010 \frac{M}{M_{\mathrm{TOV}}}.
\label{zeta}
\end{equation}
In the last equation, $\kappa_{\mathrm{eff}}^{T} = 3/18 \tilde{\Lambda}$, with $\tilde{\Lambda}$ 
being the binary's mass-weighted tidal deformability
\begin{equation}
\tilde{\Lambda} = \frac{16}{3} \frac{(M_A + 12 M_B)M_A^4 \Lambda_A + (M_B + 12 M_A)M_B^4 \Lambda_B}{(M_A+M_B)^5}.
\label{lambdat}
\end{equation}
Although $\zeta$, and therefore $c_1$, in Eq.~\eqref{zeta} is a function of the maximum mass 
allowed for a non-rotating stable NS $M_{\mathrm{TOV}}$, which depends on the specific EoS, 
we fix $M_{\mathrm{TOV}} = 2 M_{\odot}$ for the model version with quasi-universal relations in this work \footnote{In principle we could treat $M_{\mathrm{TOV}}$ as a free parameter, but this would impair the main benefit of this version of the model, namely to avoid additional parameters to sample over. However, in the future, given the increasing number of multi-messenger detections of binary neutron stars mergers and the possibility to observe high mass pulsars \cite{Margalit:2017dij,Rezzolla:2017aly,Antoniadis:2013pzd,NANOGrav:2019jur}, one can expect to have a significantly smaller uncertainty in $M_{\mathrm{TOV}}$ than today. The value of the maximum supported mass estimated from this new information will then provide the fixed value of $M_{\mathrm{TOV}}$ to employ in our model.}. The median relative error introduced on $\zeta$ by this approximation is 0.31, for the hybrid waveforms in the SACRA and CoRe database. This error propagates to the $c_1$ parameter causing a median relative error of approximately $5\%$.  %this work. %We compute, for hybrids in both the SACRA and CoRe database, the value of $\zeta$ obtained with this assumption, and the one that is given by the correct $M_{\mathrm{TOV}}$ of the specific EoS, and we find that, on average, this approximations introduces a $\sim 12 \%$ error\footnote{The error reaches xxx for some hybrids with 125H EoS, because the real and approximated $\zeta$ have opposite signs}.
%, which on average introduces a $\sim 12 \%$ error.
%\red{Where does the 12\% come from?}

With this choice for $c_0$ and $c_1$, a model for $c_{2}$ is built from a set of 
48 non-spinning NR waveforms, from the CoRe database~\cite{Dietrich:2018phi, core_webpage}. 
For this, we first find the values of $c_{2}$ that minimize the mismatch of the 
Lorentzian waveform and the NR waveform between 0.75 $c_{1}$ and 8192 Hz 
using a flat noise power spectral density (PSD); see Sec.~\ref{sec:mismatch} for details. 
The flat PSD ensures that no high-frequency 
information is suppressed in the match computation. For each waveform,  
$c_{2}$ minimization is performed using the \texttt{`L-BFGS-B'}, 
\texttt{`SLSQP'}, \texttt{`TNC'} and \texttt{`Powell'} methods available in 
\texttt{SciPy}~\cite{2020SciPy-NMeth} and the value of $c_2$ with the least mismatch 
value is used. It was seen that $c_2$ showed a similar trend against $\kappa^T_{\mathrm{eff}}q^{2}$, 
with $q = M_A/M_B$ the mass ratio, as $c_1$ does against $\zeta$. Hence, a analogous ansatz was 
used to perform a fit. However, using the parameters obtained from doing a simple curve fit 
showed unphysical amplitude behaviour for a few of the NR waveforms. For further tuning, the 
mismatch was minimized for all the NR waveforms by varying the fit parameters and the parameters 
that gave the least mismatch were then recorded and added to the model. The functional form of 
$c_2$ and the values obtained for the fit parameters in this manner are

%Fig.??? shows that \anna{if we want the fit plot} $c_2$ plotted against $\kappa^T_{\mathrm{eff}}q^{2}$, with $q = M_A/M_B$ the mass ratio, follows a similar trend as $c_1$ does against $\zeta$. Hence, using an analogous ansatz, we find the following relation:

\begin{equation}
c_2 = 2 + \gamma \frac{1 + C \kappa^T_{\mathrm{eff}} q^2}{1 + D \kappa^T_{\mathrm{eff}} q^2},
\end{equation}
with $\gamma = 19.4579017$, $C = -9.63390738 \times 10^{-4}$, and $D = 6.45926154 \times 10^{-5}$. The median relative error for $c_{2}$ during minimization is $0.56$.

\subsubsection{The full waveform}

To obtain a model describing the full coalescence, the previously derived postmerger model is 
connected to the waveform describing the inspiral and merger part of the signal, for which 
we use the phenomenological waveform \texttt{IMRPhenomD$\_$NRTidalv2} \cite{Dietrich:2019kaq}. \\

\textbf{Amplitude:} To ensure a smooth transition\footnote{We note that the 
employed approach neglects any contribution of the postmerger signal towards frequencies below the merger frequency.} between the two models, we apply a Planck-taper window $\alpha_{\rm Pl}$:
\begin{equation}
\alpha_{\rm Pl} = \begin{cases} 0 &\mbox{for   } f < f_{\mathrm{tr}},
		\\ \exp[\frac{f_{\mathrm{end}} - f_{\mathrm{tr}}}{f - f_{\mathrm{tr}}} + \frac{f_{\mathrm{end}} - f_{\mathrm{tr}}}{f - f_{\mathrm{end}}} + 1]^{-1} &\mbox{for   } f_{\mathrm{tr}} < f < f_{\mathrm{end}},
		\\ 1 &\mbox{for   } f > f_{\mathrm{end}}.
\end{cases}
\end{equation}
The window is applied just before the frequency of the main postmerger peak $f_2$, 
which corresponds to our model's parameter $c_1$. The value of the window's starting frequency 
$f_{\mathrm{tr}}$ is chosen to ensure a good match with 
NR waveforms. In particular, in Ref.~\cite{Tsang:2019esi} one of the time-domain features 
identified in the postmerger signal morphology is the \textit{first postmerger minimum}, which 
corresponds to a clear amplitude minimum present shortly after the merger, before the amplitude 
starts increasing again. By comparison with NR waveforms in the CoRe database 
\cite{Dietrich:2018phi, core_webpage}, we found that this feature is best reproduced by our 
model when the Planck window is applied between $f_{\mathrm{tr}} = 0.75 \, c_1$ and 
$f_{\mathrm{end}} = 0.9 \, c_1$. Following \cite{Khan:2015jqa}, we add an exponential 
correction factor $\exp[-\frac{p(f-c_1)}{c_2}]$ to the Lorentzian amplitude, in order to smoothen 
possible kinks arising when going to the time domain. We set $p = 0.01$, which is enough to reduce 
the kink, but not so large that it significantly influences the merger amplitude.
%The \texttt{IMRPhenomD$\_$NRTidalv2} waveform is tapered-out starting from $3.0*f_{merger}$ \ck{should we say "waveform is extended out to 3fmerg".}, in order to have the frequency region up to our transition frequency safely covered as the merger frequency can be smaller than $0.75 f_{2}$ for some cases.
%\ck{Adding a bit here.} 
%Therefore, the amplitude $\mathcal{A}$ of the full waveform is given by
%\begin{equation}
%  \mathcal{A}(f) =  \mathcal{A}_{\rm IM}(f) + \alpha \mathcal{A}_{\rm Lor}(f) \times \exp[-\frac{p(f-c_1)}{c_2}],bility
%\end{equation}
%where $\mathcal{A}_{\rm IM}(f)$ and $\mathcal{A}_{\rm Lor}(f)$ are the amplitude of the \texttt{IMRPhenomD$\_$NRTidalv2} and Lorentzian waveform respectively.\\
%\td{I don't fully understand is $\mathcal{A}(f)$ equal to $c_0$.}

\textbf{Phase:} To ensure that the waveform phase is $C^1$ continuous, we introduce two coefficients $a$ and $b$, writing the phase as
\begin{equation}
  \phi_{\rm IM}(f) = \phi_{\rm Lor}(f) + a + bf,
\end{equation}
with $\phi_{\rm IM}$ the phase of \texttt{IMRPhenomD$\_$NRTidalv2} waveform and $\phi_{\rm Lor} = \arg(h_{22}(f))$ the Lorentzian one.
 
The values of $a$ and $b$ are computed at the same transition frequency 
$f_{\mathrm{tr}} = 0.75 \, c_1$ at which we start the Planck-taper window for the amplitude, 
such that
\begin{align}
  \left . \frac{d\phi_{\rm IM}}{df} \right\rvert_{f_{\mathrm{tr}}} &= \left . \frac{d\phi_{\rm Lor}}{df} \right\rvert_{f_{\mathrm{tr}}} + b, \\
  \phi_{\rm IM} (f_{\mathrm{tr}}) &= \phi_{Lor}(f_{\mathrm{tr}}) + b f_{\mathrm{tr}} + a.
\end{align}

Finally, to reduce the Lorentzian contribution to the pre-merger and merger amplitude, 
we multiply the waveform by a factor $\exp[-i 2\pi \Delta t f]$, which will induce a time shift of 
$\Delta t$ in the time-domain waveform; $\Delta t$ is computed as the time interval between the 
merger and the \textit{first postmerger minimum} described by Eq.~(2) in \cite{Tsang:2019esi}. 

%The waveform phase is finally given by
%\begin{equation}
%  \phi = \begin{cases} \phi_{\rm IM}, & \mbox{for   } f < f_{\mathrm{tr}}
%    \\ \phi_{\rm Lor} \times \exp[-i 2\pi dt f] + b f + a, & \mbox{for  } f > f_{\mathrm{tr}}.
%  \end{cases}
%\end{equation} 

The frequency-domain gravitational waveform can be written as
\begin{equation}
\tilde{h}(f) = \mathcal{A}(f) e^{i \phi(f)},
\end{equation}
with $\mathcal{A}(f)$ the amplitude and $\phi(f)$ the phase. Therefore, 
in our model the full waveform is given by:
\begin{widetext}
\begin{equation}
\tilde{h}(f) = \begin{cases} 
				\mathcal{A}_{\rm{IM}}(f) e^{i\phi_{\rm{IM}}} & \mbox{for   } f < f_{\mathrm{tr}}, \\ 
			\left(\mathcal{A}_{\rm{IM}}(f) + \alpha_{\rm Pl} \mathcal{A}_{\rm Lor}(f) e^{-\frac{p(f-c_1)}{c_2}}\right) 
			e^{i\left(\phi_{\rm Lor} + b f + a \right) -i 2\pi \Delta t f } & \mbox{for  } f > f_{\mathrm{tr}},
				\end{cases}
\end{equation} 
\end{widetext}
where $\mathcal{A}_{\rm IM}(f)$ and $\phi_{\rm IM}(f)$ are respectively the amplitude and phase of 
the \texttt{IMRPhenomD$\_$NRTidalv2} waveform, and 
$\mathcal{A}_{\rm Lor} = \abs{h_{22}(f)}$ the amplitude of the Lorentzian one. %To ensure that our model covers also the high frequencies possible for the postmerger, we extended the \texttt{IMRPhenomD$\_$NRTidalv2} waveform with a series of zeroes beyond its maximum frequency and up to $3 \cdot f_{\rm merg}$. 

\begin{figure}[t]
	\centering
	\includegraphics[width=1.\columnwidth]{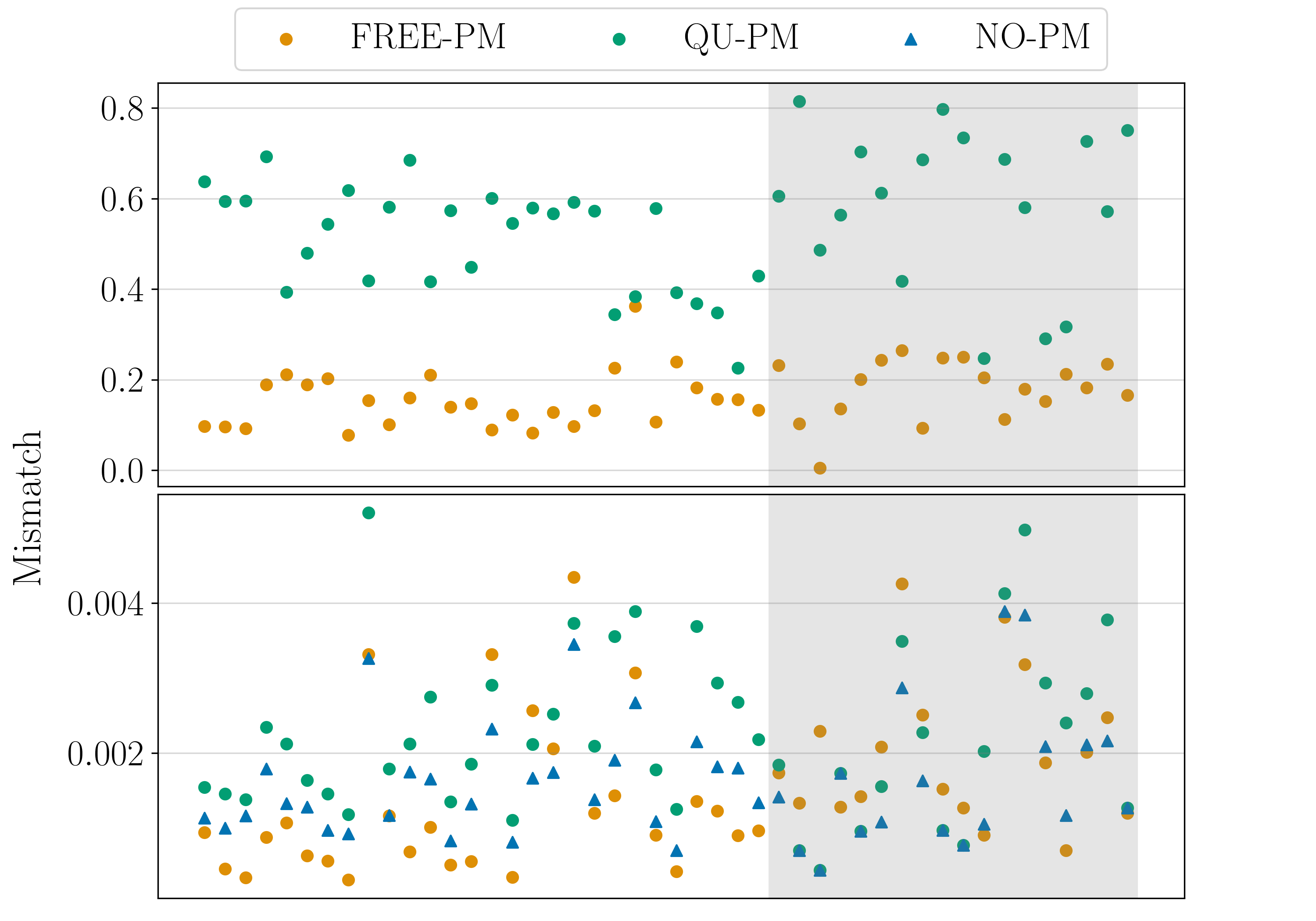} \\
	\caption{Mismatches between hybrid waveforms from the CoRe (in the gray-background band) and SACRA database, and our postmerger model, for both the versions \free{} and \qupm{}. The top panel shows mismatches in the postmerger frequency band, i.e., between $[1.1 \, f_{\rm merg}, 4096] \, \rm Hz$, the bottom panel for the whole waveform, between $[30, 4096] \, \rm Hz$. In the latter case, for comparison we show also mismatches computed between the hybrids and the \nopm{} model.}
	\label{fig:mismatch}
\end{figure}

In the following, we refer to the \texttt{IMRPhenomD\_NRTidalv2\_Lorentzian} postmerger model 
with quasi-universal relations as \qupm{}, to the one with free Lorentzian parameters as \free{}, 
and to the model without postmerger, \texttt{IMRPhenomD\_NRTidalv2}, as \nopm{}.

\subsection{Mismatch}
\label{sec:mismatch}

The mismatch between two waveforms $h_1$ and $h_2$ is defined as 
\begin{equation}
MM = 1 - \mbox{max}_{\phi_c, t_c} \frac{\left \langle h_1 (\phi_c, t_c) | h_2\right \rangle}{\sqrt{\left \langle h_1 | h_1\right \rangle \left \langle h_2 | h_2\right \rangle}},
\end{equation}
where $t_c$ and $\phi_c$ are an arbitrary time and phase shift, and the noise-weighted inner product is 
defined as
\begin{equation}
\left\langle a | b \right\rangle \equiv 4 \mbox{Re} \int_{f_{\rm low}}^{f_{\rm high}}  \frac{\tilde{a}^*(f)\tilde{b}(f)}{S_n(f)} df,
\label{eq:innerprod}
\end{equation}
where $S_n(f)$ is the noise spectral density, $\tilde{a}(f)$ the Fourier transform of $a(t)$, 
and $^*$ denotes the complex conjugate.
To validate the \texttt{IMRPhenomD\_NRTidalv2\_Lorentzian} model, we compute mismatches with 
the hybrid waveforms in the CoRe \cite{Dietrich:2018phi, core_webpage} and SACRA 
\cite{Kiuchi:2019kzt} database. The mismatch is computed with \texttt{PyCBC} 
\cite{alex_nitz_2022_5825666} functions and zero noise, i.e., with a flat PSD. For the \free{} model, to get 
the Lorentzian parameters that better describe each hybrid's postmerger, we optimize 
the mismatch over $c_1$, $c_2$; we do not include the Lorentzian maximum value $c_0$ in the 
minimization, because, giving just an amplitude scaling factor, the mismatch is insensitive to it. 
The initial values for the optimization are found with a least-squares fit on the postmerger part of 
the hybrid waveform, for $f \ge 1.3 \, f_{\rm merg}$. Fixing $c_1$ and $c_2$ to the optimal values, we then compute the optimal value for $c_0$ with a least-square fit on the hybrid's postmerger signal. 
We use the optimal values for the $c_i$ coefficients to generate the \free{} waveform, for which we compute the mismatch with the hybrid in different frequency ranges. For the \qupm{} model, instead, the Lorentzian parameters are computed from the quasi-universal relations described in Sec.~\ref{sec:parameter_models}, using the values of the hybrids'binary parameters. The top panel of Fig.~\ref{fig:mismatch} shows the mismatches in the frequency band between $[1.1 \, f_{\rm merg}, 4096] \, \rm Hz$: despite our simple description of the postmerger, when using the \free{} model for almost all hybrids mismatches lie below 0.3. Mismatches values increase systematically by roughly a factor 3 when computing them with respect to the \qupm{} model, which is expected since in this case the Lorentzian parameters are not optimized to the hybrid waveform. When considering the whole waveform, in the frequency range $[30, 4096] \, \rm Hz$, the mismatch is always below $0.005$, as shown in the bottom panel of Fig.~\ref{fig:mismatch}. Also in this case, for most hybrids the \free{} model gives better matches compared to \qupm{} one. The fact that mismatches computed over the whole waveform do not follow the trend of the ones computed only in the high frequency region is due to the fact that different values of the Lorentzian parameters translates also into different tapering and continuity conditions, influencing the late inspiral-merger phase too. For comparison, we show also the mismatches computed in the same frequency range with the \nopm{} waveform. The plot does not highlight a systematic improvement in the mismatches when using one of the models; the difference between the mismatch obtained with the \nopm{} and \free{} models varies from 0.0019 to $8\times 10^{-6}$, with an average variation of 0.0005. In some cases, the \nopm{} model gives lower mismatches than one of the models with postmerger: this occurs because the \nopm{} waveform includes no signal after the merger, therefore computing the mismatch for frequencies higher than the merger one, in a region where the waveform is zero, does not contribute to the overall mismatch, hence the lack of the postmerger signal does not reduce the match computed up to the merger frequency. However, in more than $60 \%$ of cases the mismatch is reduced when using the \free{} model, showing that our postmerger description with optimized parameters improves the signal characterization.

\subsection{Parameter estimation}

 In the following, we focus on how to recover the source's parameters given the detector data $d$ and under the hypothesis of a specific model $\mathcal{H}$ used to describe the waveform. In a Bayesian framework, this corresponds to evaluating the posterior $p(\vec{\theta}|\mathcal{H}, d)$, which, according to Bayes' theorem, is
\begin{equation}
p(\vec{\theta}|\mathcal{H}, d) = \frac{p(d|\mathcal{H}, \vec{\theta}) p (\vec{\theta}|\mathcal{H})}{p(d|\mathcal{H})} .
\label{bayes}
\end{equation}

In Eq.~(\ref{bayes}), the prior probability density $p (\vec{\theta}|\mathcal{H})$ encodes our 
prior knowledge about the source or the model; the evidence $p(d|\mathcal{H})$ describes 
the probability of observing the data $d$ given the model $\mathcal{H}$, independently of 
the specific choice of parameters $\vec{\theta}$; and the likelihood $p(d|\mathcal{H}, \vec{\theta})$ 
represents the probability of observing $d$ with the specific set of parameters $\vec{\theta}$.

The priors chosen for this work are described later in this section, while the evidence $p(d|\mathcal{H})$ serves as normalization constant of the posterior distribution, and is given by
\begin{equation}
p(d|\mathcal{H}) = \int d\vec{\theta} p(d|\mathcal{H}, \vec{\theta}) p (\vec{\theta}|\mathcal{H}) .
\label{evidence}
\end{equation}

Assuming the data $d$ consist of Gaussian noise and a GW signal $h(\vec{\theta})$, the likelihood 
can be expressed as \cite{Veitch:2009hd}
\begin{equation}
p(d|\mathcal{H}, \vec{\theta}) \propto exp \left[ -\frac{1}{2} \left\langle d - h(\vec{\theta}) | d - h(\vec{\theta}) \right\rangle \right] 
\label{likelihood}
\end{equation}
with the noise-weighted inner product defined as in Eq.~(\ref{eq:innerprod}).

%\begin{equation}
%\left\langle a | b \right\rangle \equiv 4 \mbox{Re} \int_{f_{\rm low}}^{f_{\rm high}}  %\frac{\tilde{a}^*(f)\tilde{b}(f)}{S_n(f)} df,
%\label{eq:innerprod}
%\end{equation}
%where $S_n(f)$ is the noise spectral density, $\tilde{a}(f)$ the Fourier transform of $a(t)$, 
%and $^*$ denotes the complex conjugate.

To sample the likelihood function, we use the nested sampling \cite{Skilling:2006gxv,Veitch:2009hd} 
package \texttt{dynesty} \cite{Speagle:2019ivv, sergey_koposov_2022_6456387}, which is included 
in the \texttt{bilby} library \cite{Ashton:2018jfp,Romero-Shaw:2020owr}, with 2048 live points.

\subsubsection{Relative binning}
\label{sec:relbin}

The likelihood evaluations required at each sampling step are very expensive, since, in order 
to compute the inner product, we need to evaluate the waveform on a dense and uniform frequency grid. 
The size of the grid increases both with the duration of the signal and the maximum frequency used in 
the analysis. In our case, we set $f_{\rm max} = 4096 \, \rm Hz$, since the postmerger GW signal is 
expected to lie within the few kilohertz regime. Moreover, we study BNS systems, whose low masses imply a 
long signal duration. Although we set the starting frequency to $f_{\rm low} = 30 \, \rm Hz$, the 
typical duration of the signal in band is still roughly 200 s. To overcome the issue of the computational 
cost of the analysis needed for this work, we employ the technique of relative binning 
\cite{Zackay:2018qdy, Leslie:2021ssu}, which reduces the number of waveform evaluations from 
all the points on the grid to a limited number of frequency bins.

The underlying assumption in relative binning is that the set of parameters yielding a 
non-negligible contribution to the posterior probability produce similar waveforms, 
such that their ratio varies smoothly in the frequency domain. In each frequency bin $b = [f_{\rm min}(b), f_{\rm max}(b)]$, if we choose a reference waveform $h_0(f)$ that describes sufficiently well the data, the ratio with the sampled waveforms can be approximated with a linear interpolation
\begin{equation}
r = \frac{h(f)}{h_0(f)} = r_0(h,b) + r_1(h,b)(f-f_{\rm m}(b)) + \mathcal{O}[(f-f_{\rm m}(b))^2],
\label{eq:wfm_ratio}
\end{equation}
with $f_{\rm m}(b)$ the central frequency of the bin $b$.

This allows to approximate the likelihood inner product as
\begin{equation}
	\left\langle d(f)|h(f)\right\rangle \approx \sum_{b} \left( A_0(b) r_0^*(h,b) + A_1(b) r^*_1(h,b)\right),
\end{equation}
where the summary data
\begin{align}
A_0(b) &= 4 \sum_{f \in b} \frac{d(f) h_0^*(f)}{S_n(f)/T}, \\
A_1(b) &= 4 \sum_{f \in b} \frac{d(f) h_0^*(f)}{S_n(f)/T} (f-f_m(b)) 
\end{align}
are computed on the whole frequency grid, but only for the reference waveform. Also $\left\langle h(f)|h(f)\right\rangle$ is calculated with a similar approach. In this method, the evaluation of sampled waveforms is required only to compute the bin coefficients $r_0(h,b)$ and $r_1(h,b)$ in Eq.~\eqref{eq:wfm_ratio}.
In this paper, we follow the description and implementation of~\cite{Dai:2018dca, Zackay:2018qdy}. To use relative binning with \texttt{bilby} inference, we employ the code in~\cite{janquart:2022}.
More details about the relative binning method applied to our analysis are given in Appendix~\ref{sec:relbin_app}.
%Explain + ref bilby runs settings, dynesty, relative binning (mention also that changes the initial paramters doesn't affect the result?).

\begin{figure*}[t]
	\centering
	\includegraphics[width=\textwidth]{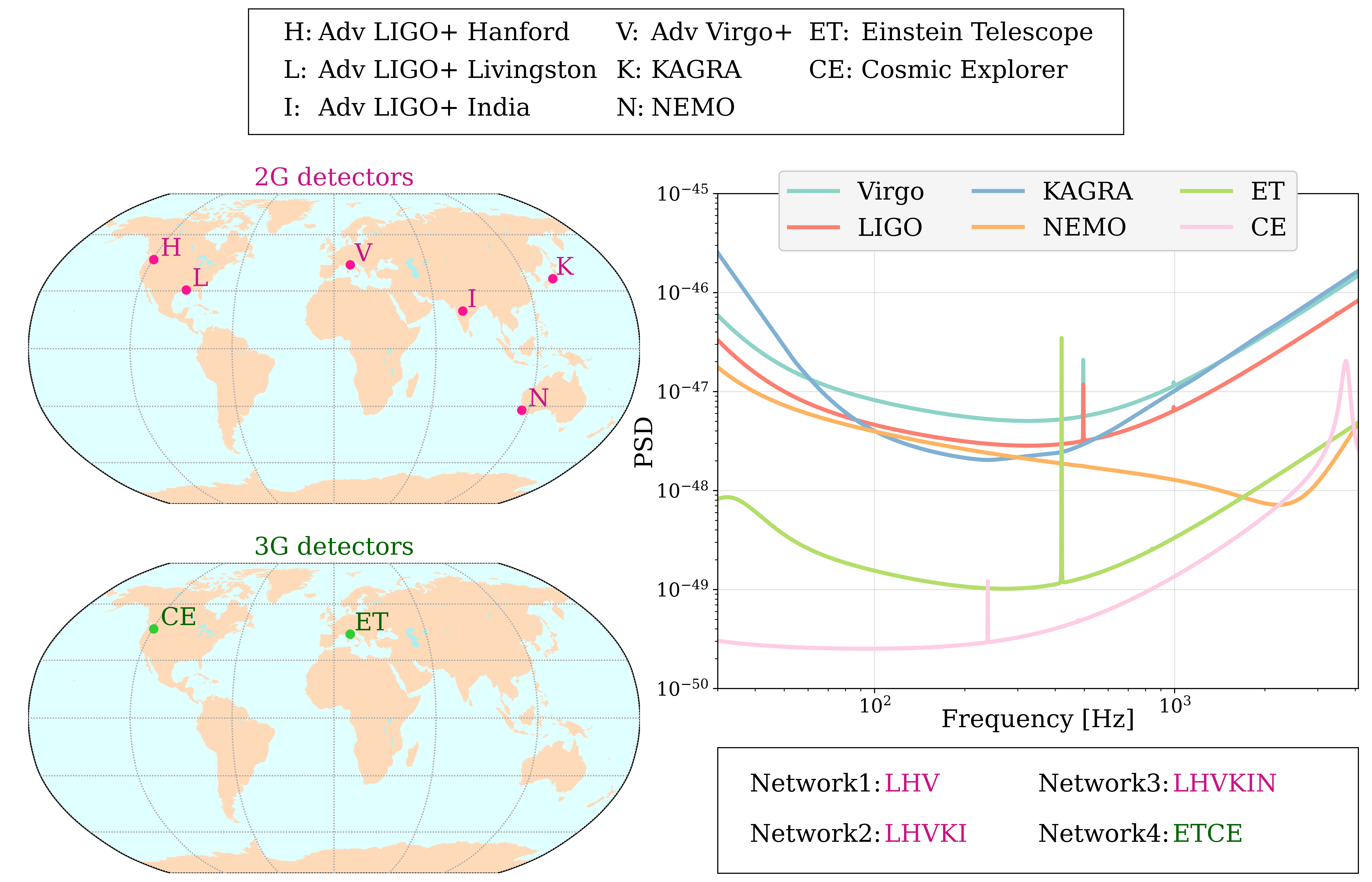} 
	\caption{Left: location of the detectors used in this study, top panel for second generation 
	(2G) detectors and bottom panel for third generation (3G) ones. Right: PSDs for the different 
	detectors. The Advanced LIGO+ PSD \cite{aLIGOplus} is used for H, L and I detectors. Since the 
	official sensitivity curve for Advanced Virgo+ is not available yet, we used the same one as for 
	the LIGO detectors, scaled by a factor 4/3 to account for the different arm-length. 
	ET sensitivity is the one referred to as 'ET-D' and given in \cite{Hild:2010id}, while CE sensitivity is given in \cite{ETCEpsd}; for KAGRA we use the PSD labeled as `Combined' 
	in  \cite{kagra_PSD}.}
	\label{fig:det_all}
\end{figure*}

\subsubsection{Simulations}

We test the performance of our model in parameter estimation analysis with simulated signals. 
We consider three different sources, and analyze them through \texttt{bilby} injections, i.e., 
using our own GW models, and through injecting NR hybrids with the same parameters; 
cf.~Tab.~\ref{tab:sources}. The employed hybrids have a postmerger signal duration of roughly 10 ms, and the postmerger contribution to their SNR for each detector network is shown in Tab.~\ref{tab:pmsnr}.

\begin{table}[t]
	\setlength\extrarowheight{2pt}
	\renewcommand{\arraystretch}{1.3}
	\begin{tabular}{ |l|c|c|c|c|c| }
		\hline
		
		Name	& $\mathcal{M}_c$ & q & $\tilde{\Lambda}$ & Injection \\ %[0.5ex]
		\hline
		Source1$_{\mathrm{[NR-inj]}}$ &  1.17524 & 0.8 & 604 &  NR: H\_121\_151\_00155 \cite{2017PhRvD..96h4060K}\\ %[0.5ex]
		Source1$_{\mathrm{[qu-pm]}}$ &  1.17524 &  0.8 & 604 &  Bilby: quasi-universal\\ %[0.5ex]
		Source1$_{\mathrm{[free-pm]}}$ &  1.17524 & 0.8& 604 & Bilby: free parameters\\ %[0.5ex]
		\hline
		Source2$_{\mathrm{[NR-inj]}}$ & 1.08819 & 1.0 & 966 &  NR: H\_125\_125\_0015 \cite{2018PhRvD..97d4044K} \\ %[0.5ex]
		Source2$_{\mathrm{[qu-pm]}}$ & 1.08819 &  1.0 & 966 &  Bilby: quasi-universal\\ %[0.5ex]
		Source2$_{\mathrm{[free-pm]}}$ & 1.08819 & 1.0& 966 & Bilby: free parameters\\ %[0.5ex]
		\hline 
		Source3$_{\mathrm{[NR-inj]}}$ &  1.17524 & 1.0 & 607 &  NR: H\_135\_135\_00155 \cite{2017PhRvD..96h4060K} \\ %[0.5ex]
		Source3$_{\mathrm{[qu-pm]}}$ &  1.17524 &  1.0 & 607 &  Bilby: quasi-universal\\ %[0.5ex]
		Source3$_{\mathrm{[free-pm]}}$ &  1.17524 & 1.0 & 607 & Bilby: free parameters\\ %[0.5ex]
		\hline
	\end{tabular}
	\caption{Properties of the sources used for injections. The NR hybrids are taken from the SACRA database \cite{Kiuchi:2019kzt}, where the employed EOSs of the NR data are simple two-piece polytropes as outlined in~\cite{Kiuchi:2019kzt}. 
	For the hybridization, we follow the procedure outlined in Sec.~III C of ~\cite{Dietrich:2018uni}. The inspiral waveform model with which we hybridize is \texttt{SEOBNRv4T}~\cite{Hinderer:2016eia}. For \texttt{bilby} injections, we used our \texttt{IMRPhenomD\_NRTidalv2\_Lorentzian} model, both with quasi-universal relations and with free Lorentzian parameters. In case of injections with the free parameters model, the injected $c_0,c_1,c_2$ values are obtained from the best fit of the correspondent NR hybrid.}
	\label{tab:sources}
\end{table}

\begin{table}[]
	\setlength\extrarowheight{2pt}
	\renewcommand{\arraystretch}{1.3}
		\begin{tabular}{l|ll|ll|ll|}
			\cline{2-7}
			& \multicolumn{2}{c}{Source1$_{\mathrm{[NR-inj]}}$}                           & \multicolumn{2}{c}{Source2$_{\mathrm{[NR-inj]}}$}                           & \multicolumn{2}{c|}{Source3$_{\mathrm{[NR-inj]}}$}                           \\ \cline{2-7} 
			\multicolumn{1}{l|}{}        & \multicolumn{1}{c|}{Total}   & \multicolumn{1}{c|}{PM}    & \multicolumn{1}{c|}{Total}   & \multicolumn{1}{c|}{PM}    & \multicolumn{1}{c|}{Total}   & \multicolumn{1}{c|}{PM}    \\ \hline
			\multicolumn{1}{|l|}{LHV}    & \multicolumn{1}{c}{100}  & \multicolumn{1}{c|}{2.0}  & \multicolumn{1}{c}{94}   & \multicolumn{1}{c|}{2.5}  & \multicolumn{1}{c}{100}  & \multicolumn{1}{c|}{2.7}  \\
			\multicolumn{1}{|l|}{LHVKI}  & \multicolumn{1}{c}{107}  & \multicolumn{1}{c|}{2.1}  & \multicolumn{1}{c}{101}  & \multicolumn{1}{c|}{2.6}  & \multicolumn{1}{c}{108}  & \multicolumn{1}{c|}{2.9}  \\
			\multicolumn{1}{|l|}{LHVKIN} & \multicolumn{1}{c}{126}  & \multicolumn{1}{c|}{6.8}  & \multicolumn{1}{c}{119}  & \multicolumn{1}{c|}{8.8}  & \multicolumn{1}{c}{126}  & \multicolumn{1}{c|}{9.9}  \\ 
			\multicolumn{1}{|l|}{ETCE}   & \multicolumn{1}{c}{1267} & \multicolumn{1}{c|}{10.2} & \multicolumn{1}{c}{1190} & \multicolumn{1}{c|}{12.3} & \multicolumn{1}{c}{1268} & \multicolumn{1}{c|}{13.3} \\  \hline
		\end{tabular}

	\caption{SNR of the NR waveforms employed in our analysis for the different detector networks (with acronyms as shown in Fig.~\ref{fig:det_all}), considering the source at a distance of 68 Mpc; we show both the SNR for the whole waveform (in the 'Total' column), computed starting at 30 Hz, and the SNR of the postmerger part of the signal (in the 'PM' column), computed starting from the merger frequency.}
	\label{tab:pmsnr}
\end{table}

%Injected signals are created from the hybrid waveforms in Table \ref{tab:hybstable}, considering different detector networks. The hybrids were chosen in order to have different SNRs in the post-merger part of the signal.
%For each injection, we recover the signal with the free parameters post-merger model, with the quasi-universal relations post-merger model and with a model without post-merger (NRTidal).

All simulated signals are injected with zero inclination $\iota$ and polarization angle $\psi$, 
and with sky location $(\alpha, \delta) = ( 0.76, -1.23) $. The sky location has been chosen such 
that none of the employed detector networks is particularly preferred.
Depending on the analysis, we performed injections at three different distances: 
225 Mpc, 135 Mpc, and 68 Mpc, which, in a network with Advanced LIGO+ and Advanced Virgo+, correspond approximately 
to a signal-to-noise ratio (SNR) of 30, 50, and 100 respectively; Table~\ref{tab:snr} reports the SNR for \stwo injections in the different detector networks and at different distances.
We take priors uniform in $[0.5,1.0]$ for mass ratio $q$, and uniform in 
$[\mathcal{M}_{c,s} - 0.05, \mathcal{M}_{c,s} + 0.05] M_\odot$ for chirp mass, where $\mathcal{M}_{c,s}$ 
is the chirp mass of the source, and the prior width is given by the precision on 
chirp-mass measurements that we anticipate for future detectors. Regarding tidal deformability 
parameters, we sample over $\tilde{\Lambda}$ and $\Delta \tilde{\Lambda}$, with a prior uniform in $[0,5000]$ and $[-5000,5000]$ respectively, where $\Delta \tilde{\Lambda}$ is defined in \cite{Wade:2014vqa} as

\begin{eqnarray} 
\Delta \tilde \Lambda &=& \frac{1}{2}\left[\sqrt{1-4\eta}(1-\frac{13272}{1319}\eta 
+ \frac{8944}{1319}\eta^2)(\Lambda_1+\Lambda_2) \right. \nonumber\\ 
&& \,\,\,\,\,\,\,+ \left.(1-\frac{15910}{1319}\eta
+\frac{32850}{1319}\eta^2+\frac{3380}{1319}\eta^3)(\Lambda_1-\Lambda_2)\right]. \nonumber\\
\end{eqnarray}

\begin{table}[]
	\setlength\extrarowheight{2pt}
	\renewcommand{\arraystretch}{1.3}
	\begin{tabular}{|c|c|c|}
		\hline
		Network & Distance [Mpc] & SNR                \\  \hline
		ETCE    & 68       & 1239 \\ 
		& 135      & 624 \\  
		& 225      & 355  \\  \hline
		LHVKIN  & 68       & 121 \\ 
		& 135      & 61   \\ 
		& 225      & 36 \\  \hline
		LHVKI   & 68       & 105 \\ 
		& 135      & 53  \\ 
		& 225      & 31 \\  \hline
		LHV     & 68       & 98  \\ 
		& 135      & 49  \\ 
		& 225      & 30 \\  \hline
	\end{tabular}
	\caption{SNR values for zero-noise \stwo{} injections in the different networks (with acronyms as shown in Fig.~\ref{fig:det_all}) and for different distances.}
	\label{tab:snr}
\end{table}
%\red{Explain $\tilde{\Lambda}_{\rm 90conf}$.}

Luminosity distance priors are uniform in comoving volume, with $D_L \in [1, 450] \rm Mpc$. 
%For coalescence phase and time we have priors uniform in $[0,2 \pi]$ and $[t_c - 0.1, t_c + 0.1]$ respectively, where $t_c$ is the trigger time of the simulation. 
Although all the sources considered are non-spinning, our baseline model \texttt{IMRPhenomD\_NRTidalv2} 
allows for aligned spins; we choose a uniform prior on the spin magnitudes $\abs{\textbf{a}_1},\abs{\textbf{a}_2} \in [0.0, 0.20]$.
%component aligned with orbital angular momentum $\abs{a_1},\abs{a_2} \in [0.0, 0.20]$.} 
Finally, when using the postmerger model with free parameters for recovery, we choose uniform priors $c_1 \in [2000,4096] \, \rm Hz$ and $c_2 \in [10,200] \, \rm Hz$, while for $c_0$ we employ a logarithmic uniform prior in $[5 \times 10^{-27}, 1 \times 10^{-22}] \, \rm s$.
%Finally, when using the free-parameters post-merger model for recovery, we choose uniform priors $c_1 \in [2000,4096] \, \rm Hz$ and $c_2 \in [10,200]$, while for the amplitude we use a logarithmic uniform prior in [5e-24,1e-22].

\subsection{Detector Networks}
\label{sec:detectors}

Earth-based GW detectors have the best sensitivity around a few tens to hundreds of Hz, which makes the inspiral and merger signal of coalescing compact objects the perfect candidate for detections. In this work, however, we are interested in the postmerger part of the signal, which is usually weaker and involves higher frequencies. Current detectors are strongly limited at these high frequencies, but the improvements planned for the future detectors' upgrades and the next generation detectors are expected to make postmerger measurements feasible. Therefore, one of the goals of this work is to assess how future detectors can improve the studies we present.
We include in our analysis the upgraded versions of existing detectors, Advanced LIGO+, Advanced Virgo+, and KAGRA, as well as new detectors whose construction has been planned for the next few years, LIGO-India and NEMO, and the next detector generation, Einstein Telescope and Cosmic Explorer.
Advanced LIGO+ design \cite{PhysRevD.91.062005} will improve the current 4 km arm-length detectors in Hanford (H) and Livingston (L) sites, including a frequency dependent light squeezing and new test masses with improved coating. Advanced Virgo+ (V), similarly, is 
the planned upgrade for the current Advanced Virgo detector in Cascina \cite{VIRGO:2014yos}. This 
transition will happen in two separate phases and include upgrades like the introduction of signal 
recycling and a higher laser power. Advanced LIGO+ and Advanced Virgo+ are the planned designs for the 
O5 observing run, which is scheduled to start roughly in 2025, and during which their BNS detection 
range will reach approximately 330 Mpc and 150-260 Mpc, respectively \cite{KAGRA:2013rdx}.
KAGRA (K)\cite{KAGRA:2020tym,Somiya:2011np, Aso:2013eba} is a 3 km arm-length interferometer 
built underground in the Kamioka mine in Japan, which already employs innovative technologies like 
cryogenic mirrors. For O5, its sensitivity at the end of the observing run is predicted to allow a 
BNS range of at least 130 Mpc \cite{KAGRA:2013rdx}.
The LIGO network involves a third detector in India (I) \cite{Saleem:2021iwi}, 
which is currently under construction and is expected to become operative approximately in 2025.
Finally, the Neutron Star Extreme Matter Observatory, or NEMO (N), is an Australian proposal 
for a gravitational-wave detector with 4 km arm-length, specifically designed to have a high sensitivity 
in the kilohertz band \cite{Ackley:2020atn}. The possible location of NEMO has not been decided yet, therefore for this work we arbitrary place it at the location shown in Fig.~\ref{fig:det_all}. Although not officially approved yet, we include it in our analysis, since its high-frequency sensitivity is particularly interesting for postmerger studies.

3G detectors are expected to increase the sensitivity by a factor between 10 and 30 
\cite{KAGRA:2013rdx} with respect to current LIGO detectors, but they require the construction of 
new facilities and are expected to start observing in the mid 2030s. At the moment, the planned 
3G detector network includes plans for Cosmic Explorer (CE) in the US and Einstein Telescope (ET) 
in Europe. CE \cite{Reitze:2019iox, Evans:2021gyd} is planned as an L-shaped interferometer 
with 40 km arm-length\footnote{Recently, also a configuration consisting of a 40 km and an 
additional 20 km detector has received attention and was considered as the reference concept 
for the recent Horizon study of~\cite{Evans:2021gyd}. In~\cite{Srivastava:2022slt}, also a tunable design for the CE detector was proposed, which would enhance sensitivity in the kilohertz band.}. For the purpose of this paper, we assume it placed at the current Hanford site. ET design \cite{Punturo:2010zz, Maggiore:2019uih}  
includes a so-called `xylophone' configuration, which guarantees an improved sensitivity at high and 
low frequencies at the same time \cite{Hild:2010id}. The two candidates for the ET site are 
Sardinia, in Italy, and Limburg, at the border between the Netherlands, Germany, and 
Belgium\footnote{In addition, recent interest arose for a third possible site located in the 
eastern part of Germany.}. For this work, we assume ET is placed at the current Virgo site. 
Although the final design of ET is still under development, here we consider it as a triangular 
detector, i.e., composed of three V-shaped interferometers with a 60 degree opening angle and 10 km arms.

In this work, we study four different detector networks: 
HLV, HLVKI, HLVKIN, and ETCE. The detectors' locations and sensitivities are shown in 
Fig.~\ref{fig:det_all}.

\begin{figure*}[t]
	\centering
	\includegraphics[width=1\textwidth]{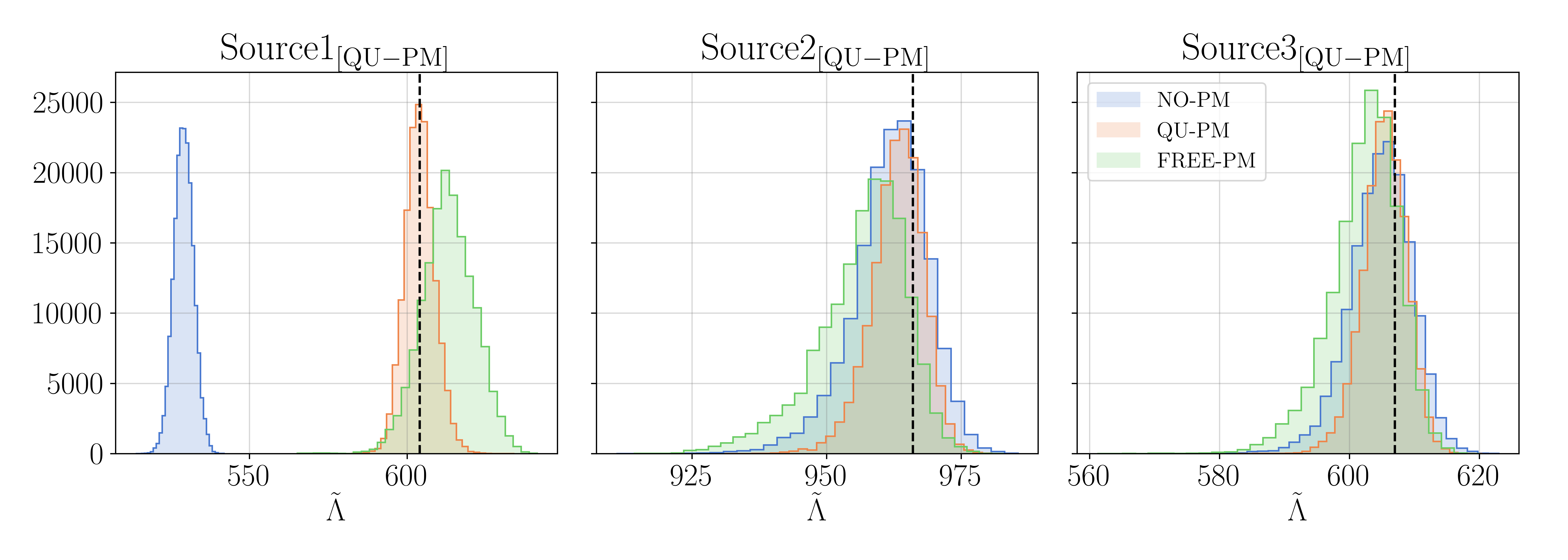} \\
	\caption{ Posterior probability density for $\tilde{\Lambda}$ in the case of \texttt{bilby} 
	injections with the \qupm{} model, for sources at 68 Mpc and with the ETCE network, 
	and recovery with the three different models \nopm{}, \qupm{} and \free{}, in blue, orange and green 
	respectively. The black dashed lines correspond to the injected values.}
	\label{fig:lambda_bcs}
\end{figure*}

\begin{figure}[t]
	\centering
	\includegraphics[width=0.98\columnwidth]{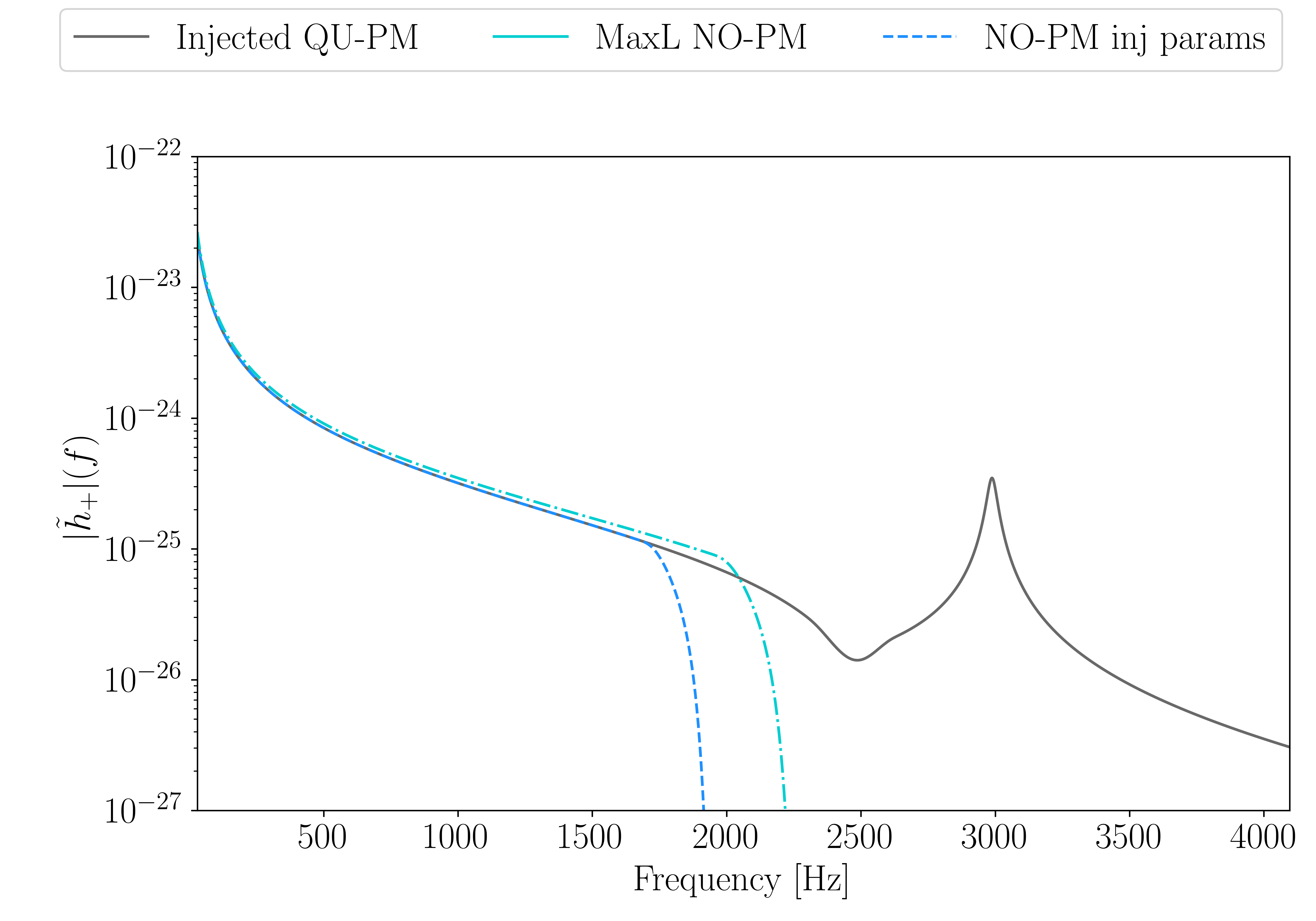} \\
	\caption{Injected signal for \sone (gray solid line), compared to the \nopm{} waveform generated with the injected parameters (dashed blue line) and with the maximum likelihood parameters recovered with the \nopm{} model (dash-dotted cyan line).}
	\label{fig:nopm_rec}
\end{figure}

\begin{figure*}[t]
	\centering
	\includegraphics[width=1\textwidth]{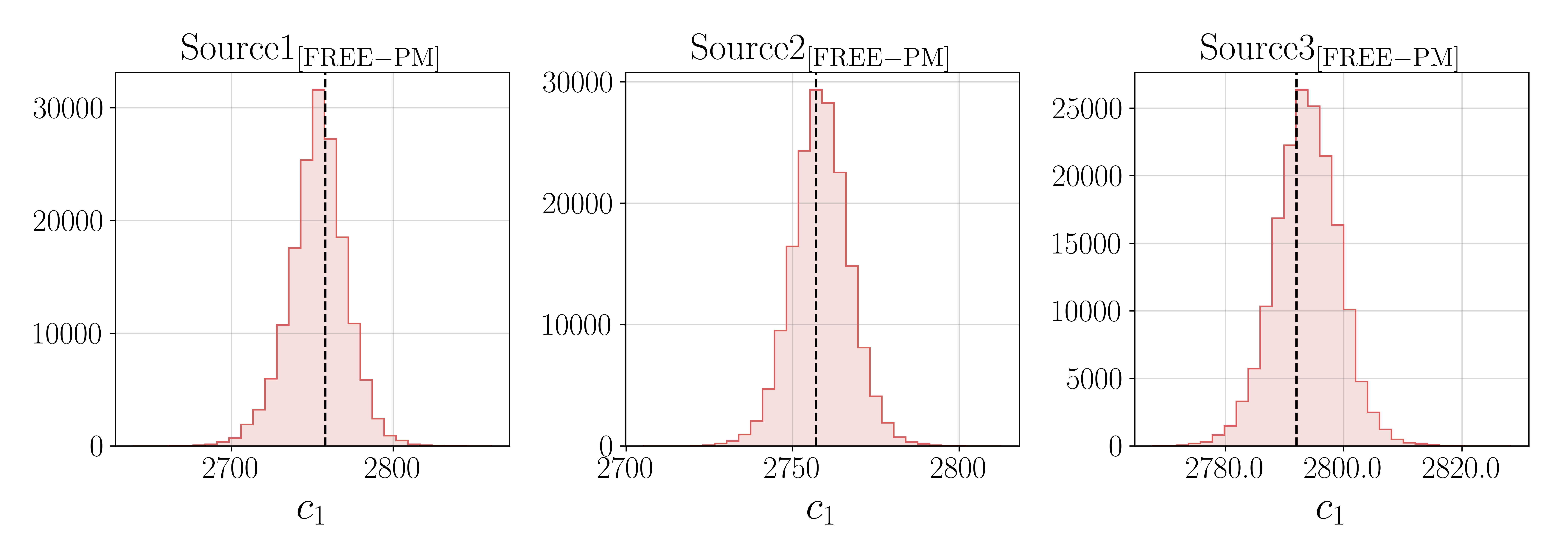} \\
	\caption{Posteriors of $c_1$ parameters for the three different sources, obtained when using the \free{} model both for injection and recovery. The black dashed lines show the injected values.}
	\label{fig:free_bcs}
\end{figure*}

\section{Results}
\label{sec:results}

In the following, we present the results of our simulations, for what concerns both the performance of 
our model and the improvement we obtain with future detectors. When using the postmerger model 
with quasi-universal relations, we are mainly interested in studying how well we can recover the 
tidal deformability parameter $\tilde{\Lambda}$. Since the quasi-universal relations that we derived 
depend on $\tilde{\Lambda}$, we expect that the postmerger part of the signal, when detected, 
brings additional information about this parameter. This will likely lead to a narrower 
posterior with respect to what we can obtain using a model without postmerger. In the case of the 
postmerger model with free Lorentzian parameters, we study how well the Lorentzian parameters 
$c_0, c_1, c_2$ can be recovered, and especially $c_1$, since it represents the frequency of the main 
postmerger emission peak.
%In the following, we refer to the \texttt{IMRPhenomD\_NRTidalv2\_Lorentzian} postmerger model 
%with quasi-universal relations as \qupm{}, to the one with free Lorentzian parameters as \free{}, 
%and to the model without postmerger, \texttt{IMRPhenomD\_NRTidalv2}, as \nopm{}. 

\subsection{Best-case scenario}
\label{sec:bcs}

We start by testing both versions of our model, \free{} and \qupm{}, in the best-case scenario, 
i.e., for \texttt{bilby} injections in zero noise, for sources as described in 
Table~\ref{tab:sources}, at a distance of 68 Mpc and with ETCE network.
Figure~\ref{fig:lambda_bcs} shows the posterior probability density of $\tilde{\Lambda}$ 
for signals obtained with \qupm{} injections, and recovered with both our postmerger models, 
\qupm{} and \free{}, and with the model without postmerger \nopm{}. As expected, 
the $\tilde{\Lambda}$ posterior becomes tighter when going from the \nopm{} to \qupm{} model, 
with the width of the $90\%$ confidence interval reducing by about $30\%$, from 23.11 to 
15.84 in the case of \stwo{}, and from 15.42 to 11.07 for \sth{}. In the \free{} recovery case, 
the posteriors become wider, with the width of the $90\%$ confidence interval reaching 27.66 for \stwo.
% and 32.12 for \stwo{} and \sth{} respectively. 
We also note that when recovering with this model, the median of $\tilde{\Lambda}$ is slightly 
underestimated with the respect to the injected values. Both these features are predictable due to 
the higher number of parameters we have to sample over. For \sone, the injected value lies 
outside the \nopm{} $\tilde{\Lambda}$ posterior distribution, but is well recovered with both the 
\qupm{} and \free{} models. Given that the sampler converged to the maximum likelihood values 
for the parameters, this shift is not caused by sampling issues, but is probably due to the fact 
that injections are performed with a signal with postmerger, and when we recover with a model without 
the postmerger description, the waveform tries to latch on to the signal after the merger, causing a 
bias in the parameter estimation. This is confirmed by the comparison, shown in Fig.\ref{fig:nopm_rec}, 
between the injected \stwo{} waveform, the \nopm{} waveform generated with the maximum likelihood 
parameters recovered with the \nopm{} model, and the one generated with the injected parameters. 
The maximum likelihood \nopm{} waveform tries to recover part of the injected postmerger signal, 
%with in the merger happening at higher a frequency with respect to what the \nopm{} model would predict with the injected parameters, and 
resulting in a deviation 
%of the maximum likelihood \nopm{} waveform 
with respect to the \nopm{} waveform obtained from the injection parameters, which explains the bias in the 
$\tilde{\Lambda}$ posterior.

Figure~\ref{fig:free_bcs} shows the posteriors for the $c_1$ Lorentzian parameter in the case of 
injection and recovery with the \free{} model, for the three different sources. 
The injected values of $c_0, c_1, c_2$ are the ones that give the best fit on the NR hybrid 
with the same binary parameters of the source considered. The $c_1$ parameter, which corresponds to 
the frequency of the main postmerger emission peak, is well recovered in all cases. 
Although we are mainly interested in the recovery of $c_1$, the \free{} model provides 
posteriors also for the $c_0$ and $c_2$ parameters, which are related to the maximum 
amplitude and width of the Lorentzian respectively. Note that the $c_0$ and $c_2$ parameters, 
which are not shown in the figure, are not recovered as well as the $c_1$ parameter, but their injected values lie in the posteriors $90\%$ confidence interval in all cases,
% but they are close to the injected values, 
as reported in Table \ref{tab:freepar}. While 
our model works for our main purpose of measuring the frequency of the dominant postmerger peak, 
the shifts that we see in the other parameters suggest that we can further improve the \free{} model; 
see e.g.~\cite{Wijngaarden:2022sah,Breschi:2022xnc} for recent developments including 
postmerger features beyond the main emission frequency. 

\begin{table}[tbh]
	\setlength\extrarowheight{5pt}
	\begin{tabular}{ |l|c|c|c|c|c| }
		\hline
		
			& $\log c_0$ & $\log c_{0,\mathrm{inj}}$ & $c_2$ & $c_{2,\mathrm{inj}}$ \\ [0.5ex]
		\hline
		
		Source1$_{\mathrm{[free-pm]}}$ &  $-56.79^{+ 0.29}_{-0.34}$ & -56.65 & $96.40^{+52.55}_{-36.90}$ & 74.0 \\ [0.5ex]
		\hline
		
		Source2$_{\mathrm{[free-pm]}}$ & $-56.18^{+0.21}_{-0.24}$ & -56.15 & $52.01^{+19.13}_{-14.19}$ & 48.0 \\ [0.5ex]
		\hline 
		
		Source3$_{\mathrm{[free-pm]}}$ & $-55.89^{+0.19}_{-0.211}$ & -55.90 & $41.26^{+14.06}_{-9.52}$ & 39.0 \\ [0.5ex]
		\hline
	\end{tabular}
	\caption{Median with $5\%$ and $95\%$ quantile values of the posterior probability density for 
	the $c_0$ and $c_2$ parameters, together with their injected values, for each of the three sources 
	analyzed, in the case of injection and recovery with the \free{} model.}
	\label{tab:freepar}
\end{table}

\subsection{Detector network performances in zero-noise}
\label{sec:zeronoise}

\begin{figure}[t]
	\centering
	\includegraphics[width=1\columnwidth]{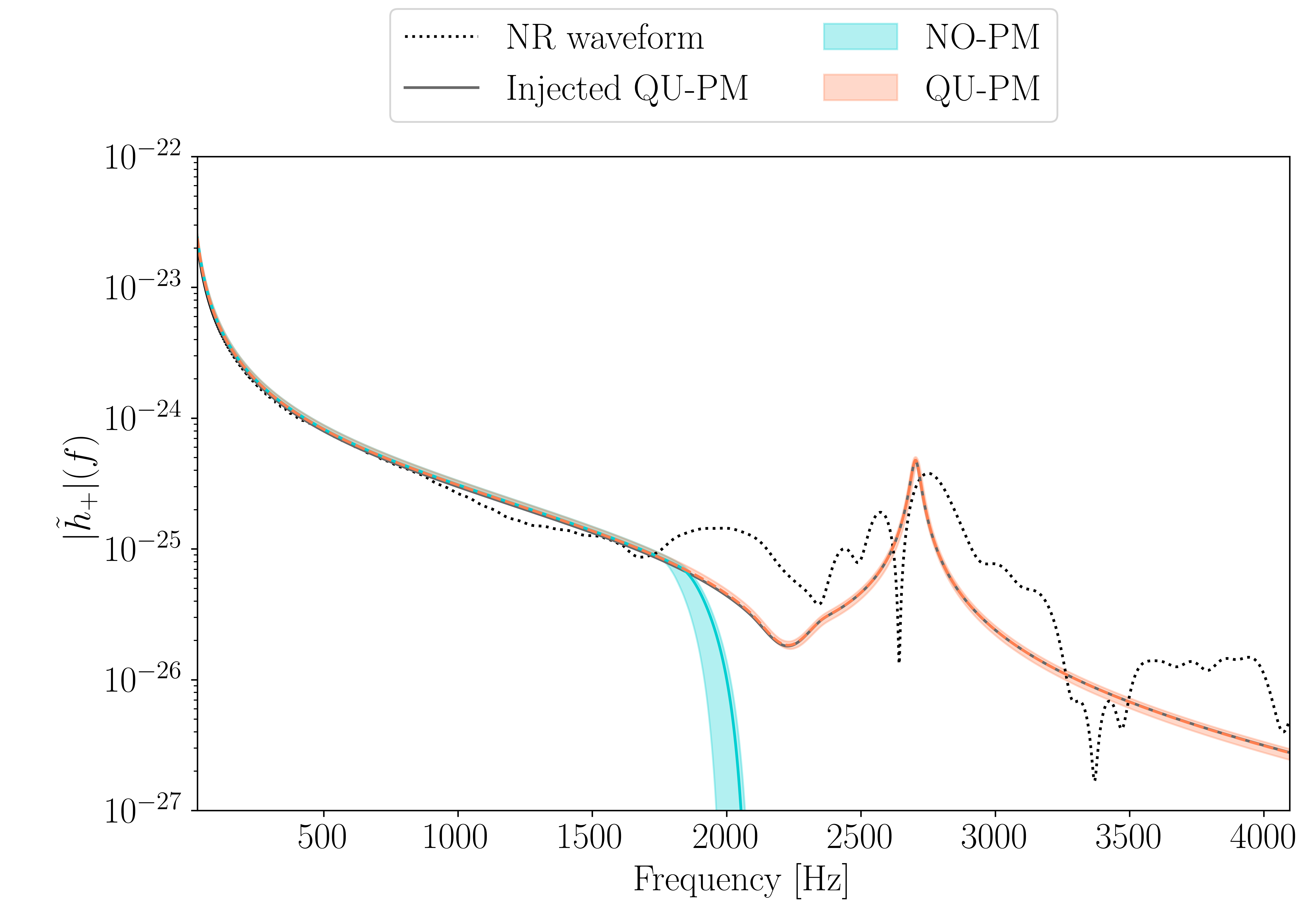} 
	\caption{Frequency domain waveform for \stwo{}, injected at a distance of 68Mpc and using the \qupm{} model (gray solid line), and the corresponding NR waveform (black dotted line). The dashed orange line and the cyan solid line show the maximum likelihood waveforms recovered for a zero noise injection in the ETCE network with the \qupm{} and \nopm{} model respectively. The orange and cyan shaded regions show the $90 \%$ confidence interval on the recovered waveforms for the two models.}
	\label{fig:secB_inj_rec}
\end{figure}

\begin{figure}[t]
	\centering
	\includegraphics[width=1\columnwidth]{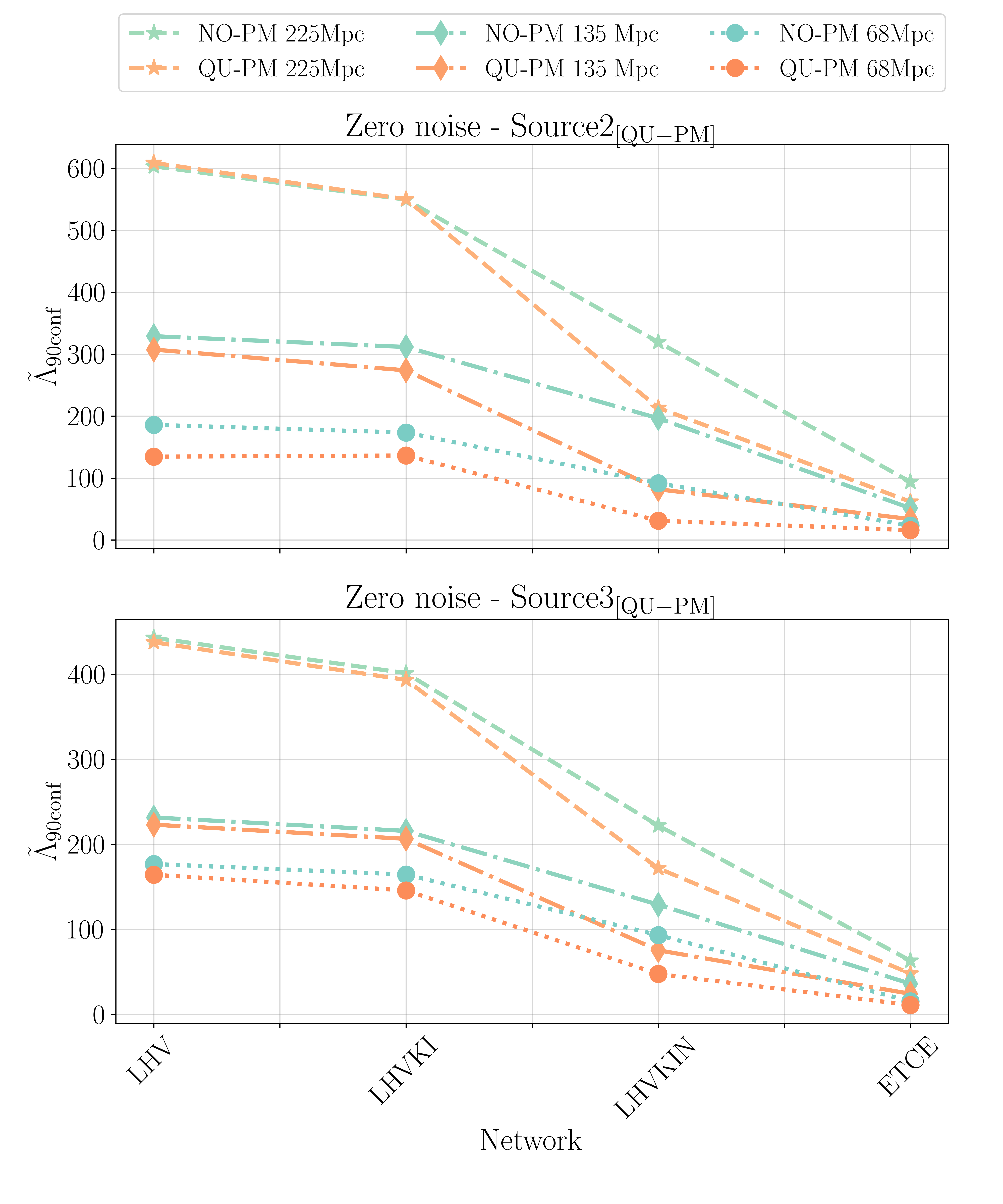} 
	\caption{Width of the 90$\%$ confidence interval of $\tilde{\Lambda}$ posterior for \stwo{} 
		(top panel) and \sth{} (bottom panel), as function of the different detector networks. 
		Orange shades represent recovery with the \qupm{} model, green shades with the \nopm{} one.}
	\label{fig:zeronoise_dlambda}
\end{figure}

We want to investigate how future detector networks will improve our postmerger analysis. 
For this purpose, we inject signals obtained from the \qupm{} model in zero noise, and recover 
both with the \qupm{} and the \nopm{} model. We analyze signals injected at three different 
distances (68 Mpc, 135 Mpc, and 225 Mpc), and we compare results for the four detector networks 
LHV, LHVKI, LHVKIN, and ETCE (as described in Sec.~\ref{sec:detectors}).
Due to limited computational resources, we look only at two different sources, 
\stwo{} and \sth{}. \\

Figure~\ref{fig:secB_inj_rec} shows the \stwo{} injected signal, and the correspondent NR waveform: the signal injected with our \qupm{} model describes well the main postmerger emission peak, but the NR waveform morphology includes also different sub-dominant emission peaks which our single Lorentzian cannot describe, and more structure in the frequency region right after the merger. Both these features should be addressed in future improvements of the model. In the same figure we show the maximum likelihood waveforms recovered both with the \qupm{} and the \nopm{} model, for a zero noise injection with the ETCE network. The recovered maximum-likelihood \qupm{} signal overlaps with the injected one, showing how well 3G detectors will be able to recover this kind of signals. In the inspiral region, this applies also to the \nopm{} maximum likelihood waveform. The inspiral signal, which we see is well recovered also with the \nopm{} model, already contains information about the $\tilde{\Lambda}$ parameter; therefore, for a ETCE network with such high SNR, we expect that little contribution to $\tilde{\Lambda}$ measurement comes from the postmerger part of the signal, given that this parameter is already very well constrained from the inspiral.

Figure~\ref{fig:zeronoise_dlambda} shows the uncertainty 
$\tilde{\Lambda}_{\rm 90conf}$, computed as the width of the 90$\%$ confidence interval of the $\tilde{\Lambda}$ posterior probability density, as a function of the detector network employed for the analysis, 
comparing the different distances and recovery models.
As expected, Fig.~\ref{fig:zeronoise_dlambda} shows that for all the detector networks considered, 
and for both models, the width of the $90\%$ confidence interval decreases with decreasing 
distance. In particular, for an LHV network, we find an improvement of 
$\sim 50\%$ when going from 225 Mpc to 135 Mpc, and of $\sim 25\% $ (for \stwo{} even $56 \%$)  
when going from 135Mpc to 68 Mpc, for both models; for the ETCE network we find 
an improvement $\sim 45\%$ when going from 225Mpc to 135 Mpc, and $\sim 55\%$ when going from 
135 Mpc to 68Mpc. Using the \qupm{} model yields systematically tighter constraints 
on $\tilde{\Lambda}$, thanks to the additional information arising from the 
quasi-universal relations that describe the postmerger part of the signal. 
For both the sources, in the case of injections at 225 Mpc and with the LHV or LHVKI network, 
we see no significant differences in $\tilde{\Lambda}_{\rm 90conf}$ in the case of 
recovery with the \qupm{} or \nopm{} model. Considering that such injections generate an 
$\rm SNR \simeq 30$ in the case of LHV network, this is consistent with the fact that in 
these situations we do not detect the postmerger signal.\\
Interestingly, the best improvement when using the \qupm{} model comes in the case of LHVKIN network. 
Going from the LHVKIN to the ETCE network, the constrain on $\tilde{\Lambda}$ improves of about 
$\sim 70 \%$ for both models, while adding NEMO to the LHVKI network leads to an improvement in 
$\tilde{\Lambda}_{\rm 90conf}$ of $\sim 60\%$ for the \qupm{} model, against the just 
$\sim 40\%$ for the \nopm{} one.
For both sources, we also see that for the LHVKIN network the constraint on $\tilde{\Lambda}$ 
obtained with the \qupm{} model for injections at 135 Mpc is better than the one we retrieve with 
the \nopm{} model for injections at 68 Mpc. 3G detectors are expected to have the 
best sensitivity over the whole frequency band, and indeed we see that for the ETCE network we 
get the smallest $\tilde{\Lambda}_{\rm 90conf}$ for both models. However, the high sensitivity at 
lower frequencies allows to obtain precise measurements of $\tilde{\Lambda}$ from the inspiral part of the signal alone,
%a very narrow posterior density distribution for $\tilde{\Lambda}$ from the inspiral part of the signal alone, 
therefore reducing the impact 
of the possible information gained from postmerger. In the case of LHVKIN network, instead, 
the constraint on $\tilde{\Lambda}$ from the inspiral is the one of second-generation detectors, 
but the high sensitivity of NEMO in the kilohertz band leads to a better detection of the postmerger, 
and therefore to significantly tighter constraints when using the \qupm{} model. If its realization 
is approved, adding NEMO to the network of second-generation detectors will significantly help the 
detection of postmerger signals and related studies. We note that for this work we analyze signals 
with a lower frequency cutoff $f_{\rm low} = 30$ Hz, missing many inspiral cycles;   
in reality, an additional improvement on $\tilde{\Lambda}$ measurements will be 
provided by the use of a lower $f_{\rm low}$.

\begin{table}[tbh]
	\setlength\extrarowheight{5pt}
	\begin{tabular}{ |l|c|c|c|c| }
		\hline
		
		& Model & $\tilde{\Lambda}_m$ $\mathrm{noise}_{\rm A}$ & $\tilde{\Lambda}_m$ $\mathrm{noise}_{\rm B}$ & $\tilde{\Lambda}_{\mathrm{inj}}$ \\ [0.5ex]
		\hline
		\stwo{} & \qupm{} & $956.68_{-8.37}^{+7.08}$  & $959.93_{-8.71}^{+6.87}$ & 966 \\
		& \nopm{} & $966.35_{-11.82}^{+9.35}$  & $953.10_{-19.11}^{+13.11}$ & 966 \\ \hline
		\sth{} & \qupm{} & $608.04_{-6.27}^{+11.65}$ & $602.36_{-12.49}^{+7.86}$ & 607 \\
		& \nopm{} & $611.76_{-7.51}^{+6.68}$ & $604.35_{-7.70}^{+6.84}$ & 607 \\
		
		\hline
	\end{tabular}
	\caption{Median values with $90\%$ confidence interval for the posterior probability density of $\tilde{\Lambda}$ in case of two different noise realizations, labeled as $\mathrm{noise}_{\rm A}$ and $\mathrm{noise}_{\rm B}$, for injections at 68 Mpc in the ETCE network and for recovery with the two different models \qupm{} and \nopm{}; the last column reports the injected value of $\tilde{\Lambda}$.}
	\label{tab:lambda_shift}
\end{table}

\subsection{Detector Network Performances in non-zero noise}
\label{sec:noise}

\begin{figure*}[t]
	\centering
	\includegraphics[width=\textwidth]{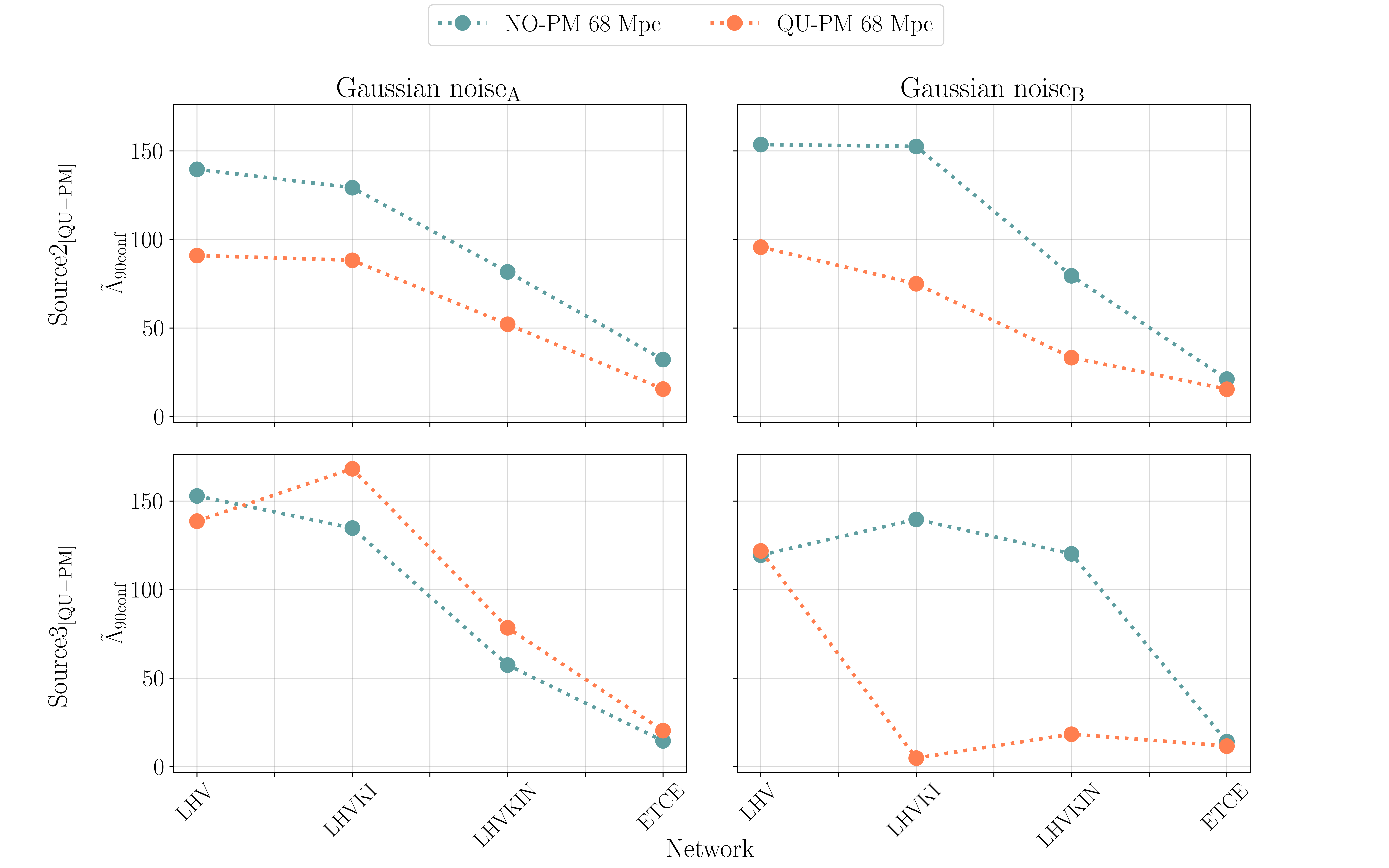} 
	\caption{Width of the 90$\%$ confidence interval of $\tilde{\Lambda}$ posterior for \stwo{} (top row) and \sth{} (bottom row), as a function of the different detector networks, obtained with two different noise realizations, $\rm noise_{\rm A}$ for the left panels, and $\rm noise_{\rm B}$ for the right ones.}
	\label{fig:gaussiannoise_dlambda}
\end{figure*}

In the previous sections we focused on model and network performances, using injections in zero noise. 
Now we want to look at the influence of noise on our study. For this reason, we repeat the analysis 
using Gaussian noise. Due to limited computational resources, we restrict to only two sources, \stwo{} 
and \sth{}, and to one distance, 68 Mpc. We inject signals using the 
\qupm{} model, and recover both with the \qupm{} and \nopm{} models, comparing results for the different 
detector networks LHV, LHVKI, LHVKIN, and ETCE. Figure~\ref{fig:gaussiannoise_dlambda} shows 
$\tilde{\Lambda}_{\rm 90conf}$ for the different detector networks. In order to assess the impact of 
noise fluctuations, we show results for two different noise realizations, which we call 
$\rm noise_{\rm A}$ and $\rm noise_{\rm B}$. Due to the noise impact on the analysis, we do not see 
the clear trends that we found in the zero noise runs, as described in the previous section 
Sec.~\ref{sec:zeronoise}. In the case of \sth{} (bottom panels in 
Fig.~\ref{fig:gaussiannoise_dlambda}), with the $\rm noise_{\rm A}$ realization the constraints 
obtained with the \qupm{} model are even wider than the ones recovered with the \nopm{} model. The most extreme fluctuation is found for \sth{}, in the case of LHVKI network and \qupm{} model, 
for which $\tilde{\Lambda}_{\rm 90conf} = 88.26$ in case of $\rm noise_{\rm A}$ and 
$\tilde{\Lambda}_{\rm 90conf} = 4.84$ for $\rm noise_{\rm B}$.
However, we see that in general $\tilde{\Lambda}_{\rm 90conf}$ decreases with more advanced 
detectors, with an improvement between $80\%$ and $90\%$ when going from the LHV to the ETCE network.
In most cases the \qupm{} model allows us to better determine $\tilde{\Lambda}$, although the 
quantitative improvement strongly depends on the source and especially on the noise realization. 
Moreover, noise fluctuations impact also the median of the $\tilde{\Lambda}$ posterior probability 
density, causing different shifts with respect to the injected values (see Table \ref{tab:lambda_shift}). 
Although such shifts appear to be small, they can cause the posterior's median to lie outside 
the $90\%$ confidence interval, especially in the case of ETCE network, where the 
$\tilde{\Lambda}_{\rm 90conf}$ is indeed very small.

% We still see the trend of $\tilde{\Lambda}_{\rm 90conf}$ decreasing when going to more advanced detector networks, but in some cases the \qupm{} model performs worse than the \nopm{} one. In order to assess the impact of noise fluctuations, we show results for two different noise realizations. \textit{(Hopefully) when using one noise realization the \qupm{} gives tighter constraints than the \nopm{} one, but for the other realization is the other way round. This shows how noise fluctuations can impact the analysis, especially when dealing with the postmerger signal.} Moreover, noise fluctuations impact also the median of the $\tilde{\Lambda}$ posterior probability density, causing some shifts with respect to the injected values (see Table \ref{tab:lambda_shift}). Although such shifts appear to be small, they can cause the posterior's median to lie outside the $90\%$ confidence interval, especially in the case of ETCE network where the $\tilde{\Lambda}_{\rm 90conf}$ is very small.

\begin{figure*}[t]
	\centering
	\includegraphics[width=1\textwidth]{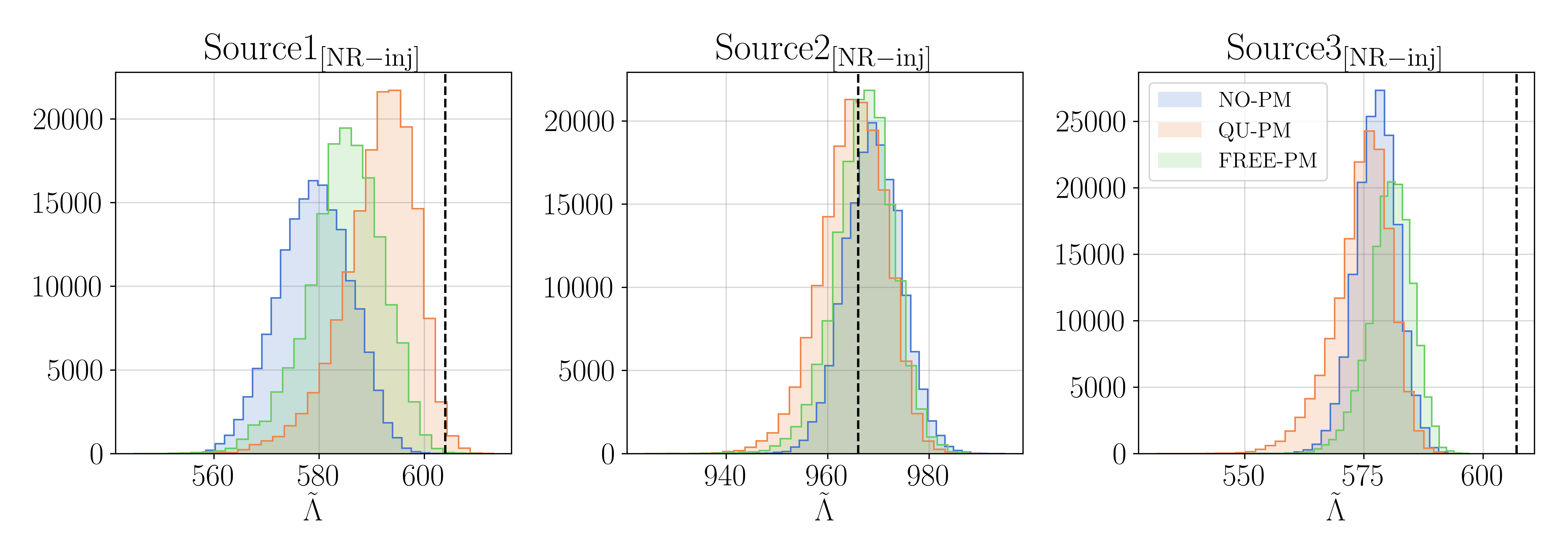} \\
	\caption{Posterior probability density for $\tilde{\Lambda}$ as recovered with the different models (\nopm{}, \qupm{} and \free{}) in the case of signals simulated by injecting NR waveforms in Gaussian noise at a distance of 68 Mpc, for the ETCE detector networks. The black dashed lines show the injected values.}
	\label{fig:hybrids}
\end{figure*}
		
\begin{figure}[t]
	\centering
	\includegraphics[width=0.98\columnwidth]{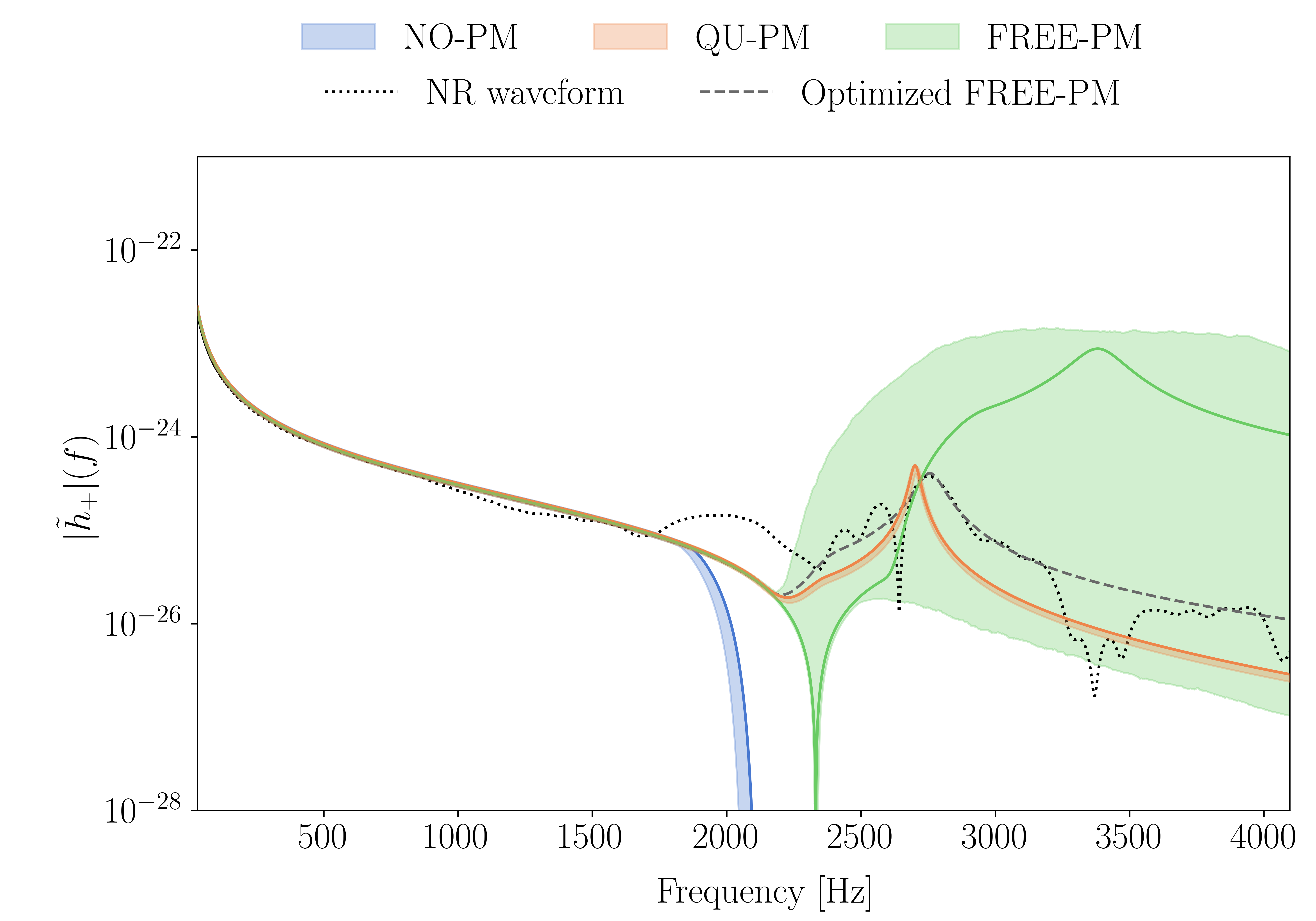} \\
	\caption{Frequency domain waveform of the injected NR (black dotted line) waveform, compared to the waveforms generated from the maximum likelihood parameters recovered for each model. The dashed gray line shows the \free{} waveform obtained by optimizing the Lorentzian parameters as explained in Sec.~\ref{sec:mismatch}. The shaded regions represent the $90 \%$ confidence interval of the posterior of the recovered waveform with the different models.}
	\label{fig:hybrids_rec}
\end{figure}

\subsection{Numerical-relativity injections}
\label{sec:nr}

Finally, we analyze simulated signals obtained by injecting NR waveforms on top of Gaussian noise. 
Figure~\ref{fig:hybrids} shows the posterior probability density of $\tilde{\Lambda}$, for injections 
at 68 Mpc in the ETCE network. For Source2$_{\mathrm{[NR-inj]}}$ the recovered posteriors of $\tilde{\Lambda}$ peak at the injected value, but for the other sources the posterior is shifted with respect to it.
For \sone{}, the $\tilde{\Lambda}$ injected value lies in the tail of the posteriors 
recovered with the \qupm{} and \free{} model, and completely outside the posterior obtained with the 
\nopm{} model; for \sth{}, the posteriors recovered with all the models peak at values between 575 and 
578, with the injected value $\tilde{\Lambda} = 607$ lying completely outside their distributions.
These shifts are due to noise fluctuations, as we showed in Sec.~\ref{sec:noise}, and possible 
limitations of our waveform models. The case analyzed here, using the ETCE network, generates a 
signal with a high SNR, and therefore a narrow posterior density for $\tilde{\Lambda}$; hence the shifts induced by noise fluctuations can result in the injected value being situated outside the $90\%$ confidence interval. Using one of the postmerger models to 
analyze signals obtained with NR waveforms does not lead to a meaningful improvement in the 
$\tilde{\Lambda}$ constraints as the ones shown in Sec.\ref{sec:bcs}. This is consistent with the fact that mismatches computed over the whole waveform (cf.lower panel of Fig.~\ref{fig:mismatch}) do not show significant improvements when using one of the postmerger models, considering that the noise and the complicated morphology of the NR injection make it more difficult for our models to recover the postmerger part of the signal, and therefore almost all the $\tilde{\Lambda}$ information comes from the inspiral. 
Nonetheless, when using the postmerger models, we see a modest improvement in the recovery of $\tilde{\Lambda}$ for Source2$_{\mathrm{[NR-inj]}}$, with respect to the \nopm{} one, and a clear improvement for Source1$_{\mathrm{[NR-inj]}}$. The latter is consistent with the results found in Sec.~\ref{sec:bcs} for \sone, where we concluded that, when using the \nopm{} model, the presence of a postmerger signal, to which the \nopm{} waveform tries to latch on, causes a bias in the $\tilde{\Lambda}$ parameter recovery.	
%for Source1$_{\mathrm{[NR-inj]}}$ we see a clear improvement in the recovery of $\tilde{\Lambda}$ 
%with the postmerger model with respect to the one without postmerger, and a modest improvement 
%is also seen for Source2$_{\mathrm{[NR-inj]}}$.}
In Sec.~\ref{sec:noise}, we saw that noise fluctuations alone can impact the performance of our 
model, but in this case an additional issue is that the NR simulations contain a more complex 
GW structure in the postmerger, which is not fully recovered with our simple Lorentzian model. This appears clearly in Figure~\ref{fig:hybrids_rec}, which shows the injected NR waveform together with the maximum likelihood ones recovered with the different models and their $90 \%$ confidence interval. The postmerger peak obtained with the \qupm{} model is slightly shifted with respect to the main postmerger peak of the NR waveform; however, the same shift was present also in the \stwo{} injected waveform in Fig.~\ref{fig:secB_inj_rec}, and hence we conclude that it is due to the imperfection of the model, not to issues in the parameter estimation process. When optimizing the mismatches to compute the best values of the fit parameters for our quasi-universal relations, it is likely that the model tries to adapt to the whole morphology of the postmerger NR signal, thus shifting with respect to what would be the description of the main emission peak only. 
For the \free{} model maximum likelihood waveform, the postmerger peak lies at a higher frequency than the true one, is much wider and with a non-physical amplitude, though this does not affect the $\tilde{\Lambda}$ recovery (cf.Fig.~\ref{fig:hybrids}). %Therefore, as concluded in Sec.~\ref{sec:bcs}, the \free{} model needs improvement for the other parameters. 
Given the large bias in $c_0$, and the fact that the injected values vary in a small range, an improvement would probably be obtained already by restricting the prior range for this parameter. For comparison, in Fig.~\ref{fig:hybrids_rec}, we show also the waveform obtained from the \free{} model with the optimized parameters computed as explained in Sec.~\ref{sec:mismatch}: the postmerger peak of the optimized \free{} waveform overlaps to the one of the NR waveform. Hence, the \free{} model can in principle describe the data well, but the additional information contained in the complex and more structured morphology of the postmerger in the hybrid signal makes it challenging for our simple model to recover all the parameters correctly.
The fact that the postmerger Lorentzian parameters cannot be recovered with a good precision causes the $90 \%$ confidence interval of the recovered waveform to be very broad. The spectra recovered with the \qupm{} model, instead, lie in a narrower interval because their values are determined by the binary's parameters, which with 3G detectors are recovered with a very high precision (see Appendix~\ref{sec:pe}). We also note that the optimized \free{} model peak does not present the same shift as the \qupm{} one, which is consistent with the fact that the mismatches in the the high-frequency region shown in Fig.~\ref{fig:mismatch} are systematically lower for the \free{} model. 
For this purpose, both our \qupm{} and \free{} models need to be improved towards more 
structured signals. Moreover, hybridization of NR waveforms starts from the few last cycles of the 
inspiral, so that also the late-inspiral and merger waveform is based on NR simulations, and thus 
different from the model we employ. The difference between the hybrids and the waveform models in the late-inspiral region is visible also in Fig.~\ref{fig:hybrids_rec}, and can lead to biases, affecting the results obtained not only with our \free{} or \qupm{} models, but also with the model without postmerger.

\section{Conclusions}
\label{sec:conclusions}

We have developed an analytical, frequency-domain model to describe the GW emission during the 
inspiral, merger, and postmerger phases of a BNS coalescence. For the inspiral and merger, we 
employed the \texttt{IMRPhenomD$\_$NRTidalv2} waveform. We incorporate the postmerger part 
through modeling the main emission peak with a Lorentzian, whose parameters, in the two versions 
of our model, are either free or determined by quasi-universal relations. 
Due to the computational cost of the analysis, our study was limited to a restricted number of BNS systems.
We have shown that in the best-case scenario of simulations with zero noise and high SNR, i.e. at a distance of 68 Mpc and with the ETCE network, the \qupm{} model leads to better constraints on the $\tilde{\Lambda}$ posteriors compared to the ones obtained with the \nopm{} model, and the \free{} model grants an accurate measurement of the frequency of the main postmerger emission peak. 
Within our study, we find that noise fluctuations can significantly impact the results; as shown in Sec.\ref{sec:noise}, they produce both large differences on the accuracy of $\tilde{\Lambda}$ measurements (quantified by the width of the $90\%$ confidence interval of the recovered $\tilde{\Lambda}$ posterior, e.g.Fig.~\ref{fig:gaussiannoise_dlambda}), and shifts in the median value of such posterior, cf. Table~\ref{tab:lambda_shift}. In some cases, this overcomes the improvement on $\tilde{\Lambda}$ measurements yielded by the use of the \qupm{} model, and calls for caution in the interpretation of the results, to distinguish the effects of a different model from the ones of noise.
 It is important to note that the shifts in $\tilde{\Lambda}$ recovery caused by noise fluctuations, which 
are evident especially in high-SNR injections, given the narrowing of the posterior, also affect the 
results obtained with the model without postmerger. 
%Therefore, they must be taken into account in parameter estimation analyses with 3G detectors, even when not related to postmerger studies.
In general, including the postmerger during the analysis provides tighter constraints on the 
$\tilde{\Lambda}$ posterior than the original inspiral-only \texttt{IMRPhenomD$\_$NRTidalv2} model. 
Finally, we used our model to recover signals obtained by injecting NR waveforms. Although we still 
see improvements in some cases when using the postmerger models, they are not as significant as 
we found for the simulated signals. This is due to noise effects and the fact that NR waveforms 
include postmerger signals with a complex structure, which a simple Lorentzian model struggles to 
recover. Despite the promising results, we conclude that our model, in both its versions, still needs 
improvements in order to be employed in the analysis of real signals.

Another central point of our study was to assess the performance of different detector networks, and 
to understand how future detectors will improve the postmerger analysis. In particular, we considered 
four different networks: (i) Advanced LIGO+ in Hanford and Livingston together with Advanced Virgo+; 
(ii) the same network as (i) extended by KAGRA and LIGO-India; (iii) the same network as (ii) 
extended with NEMO; (iv) a network consisting of a 40 km Cosmic Explorer and a 10km, triangular 
Einstein Telescope. 
Although 3G detectors, as expected, will give the best constrains on $\tilde{\Lambda}$, 
we found that NEMO, thanks to its very high sensitivity in the kilohertz band, yields the biggest 
improvement when using the \qupm{} model. 

Our study showed how, with future detector networks, GW observations from the postmerger phase of 
a BNS coalescence will allow us to unravel information about the fundamental physics describing 
supranuclear-dense matter.

\section*{Acknowledgments}

We thank Anuradha Samajdar for the useful discussion.
A.P., C.K., Y.S. and C.V.D.B. ~are supported by the research programme 
of the Netherlands Organisation for Scientific Research (NWO). This work was performed using the Computing Infrastructure of Nikhef, which is part of the research program of the Foundation for Nederlandse Wetenschappelijk Onderzoek Instituten (NWO-I), which is part of the Dutch Research Council (NWO). The authors are grateful for computational resources provided by the LIGO Laboratory and supported by the National Science Foundation Grants No.~PHY-0757058 and No.~PHY-0823459. This research has made use of data, software and/or web tools obtained from the Gravitational Wave Open Science Center (https://www.gw-openscience.org), a service of LIGO Laboratory, the 
LIGO Scientific Collaboration and the Virgo Collaboration. LIGO is funded by 
the U.S. National Science Foundation. Virgo is funded by the French Centre 
National de Recherche Scientifique (CNRS), the Italian Istituto Nazionale della 
Fisica Nucleare (INFN) and the Dutch Nikhef, with contributions by Polish and Hungarian institutes.

\appendix

%This indicates that the modifications we implemented in our model do not spoil the match with the NR hybrid waveforms. 
%We use the Lorentzian parameters obtained to generate the \texttt{IMRPhenomDNRTidalv2\_Lorentzian} waveform, for which we compute the mismatch with the hybrid in different frequency ranges. Fig.\ref{fig:mismatch} shows the mismatches between the hybrids and the \texttt{IMRPhenomDNRTidalv2\_Lorentzian} and \texttt{IMRPhenomDNRTidalv2} waveform models. The mismatches computed in the kilohertz frequency band, between $[1.1 \cdot f_{\rm merg}, 4096] \rm Hz$, show a great improvement for the \texttt{IMRPhenomDNRTidalv2\_Lorentzian} waveform, proving that our model, including the postmerger, successfully improves the match with NR waveforms. The mismatches computed over the whole waveform, in the range $[30,4096] \rm Hz$, do not show such an improvement, which is expected since the postmerger contribution becomes negligible when also the inspiral is taken into account.
\begin{figure}[t]
	\centering
	\includegraphics[width=0.98\columnwidth]{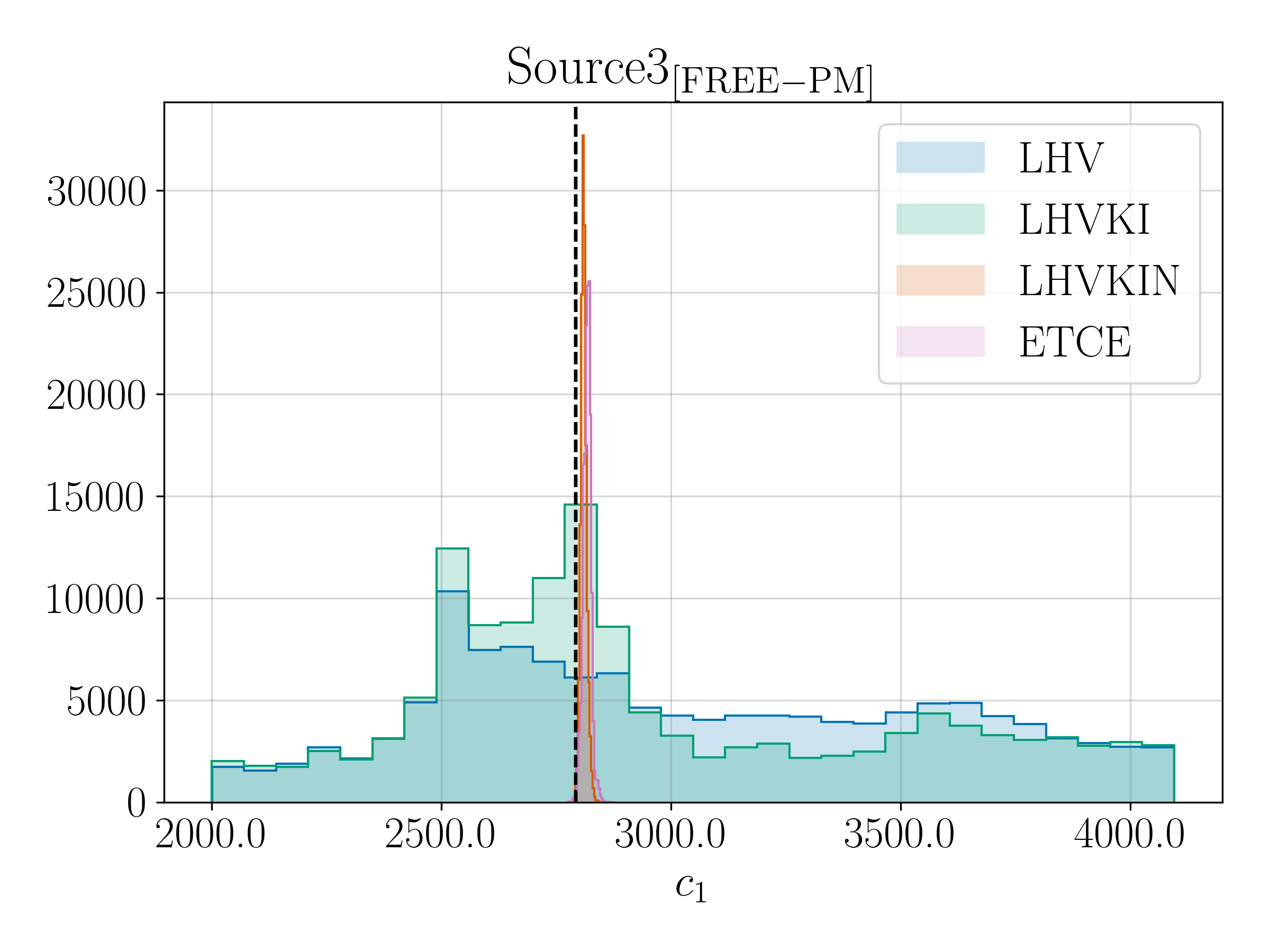}
	\caption{Posterior probability density for the $c_1$ Lorentzian parameter for the different detector networks, in the case of Gaussian noise injections at 68 Mpc. The dashed vertical line indicates the injected value.}
	\label{fig:free_det}
\end{figure}

\begin{figure*}[t]
	\centering
	\includegraphics[width=1\textwidth]{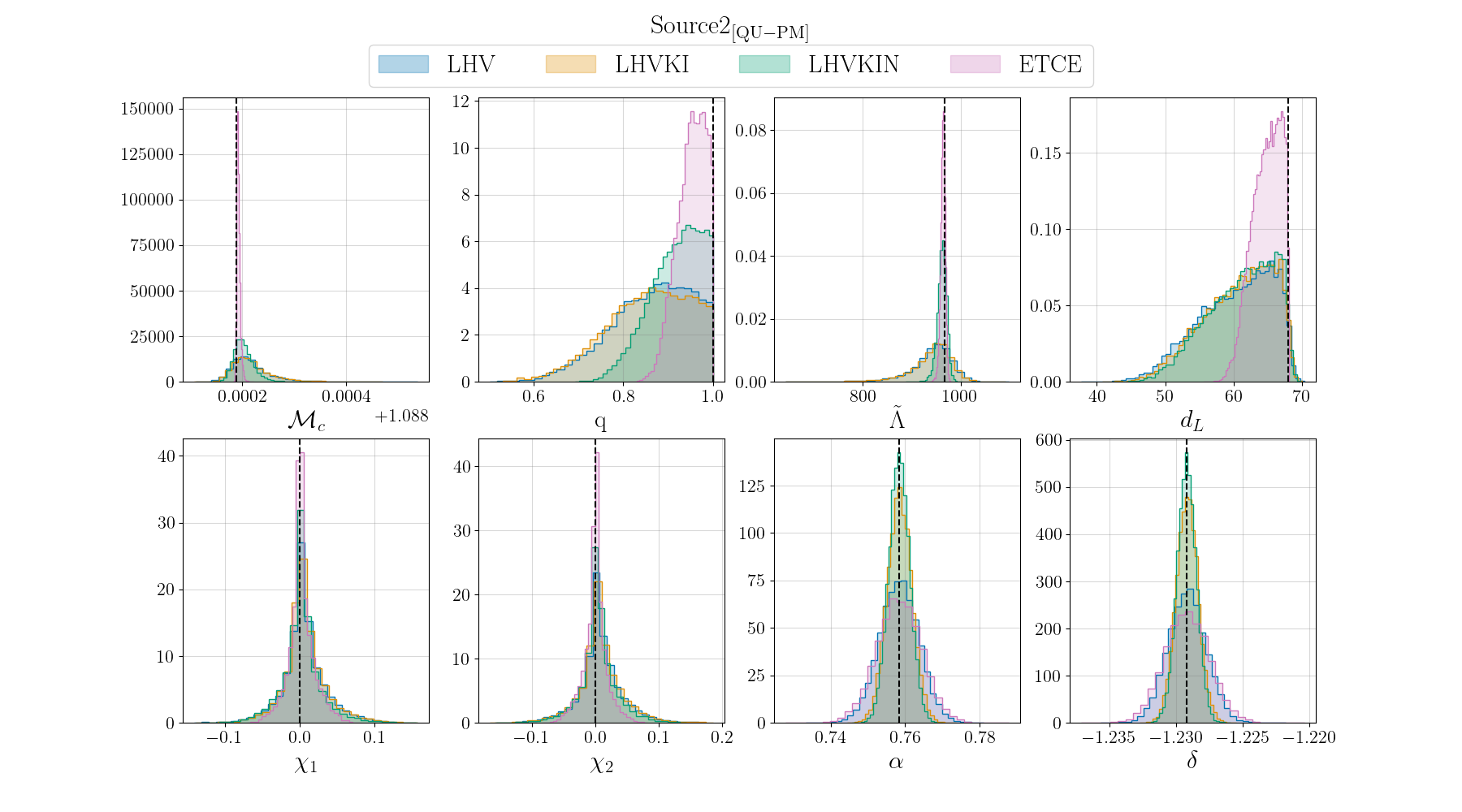}
	\caption{Normalized posterior probability density for the binary parameters recovered with the \qupm{} model with the different detector networks, for \stwo at 68Mpc; the dashed vertical lines show the injected values.}
	\label{fig:net_params}
\end{figure*}

\begin{figure*}[t]
	\centering
	\includegraphics[width=1\textwidth]{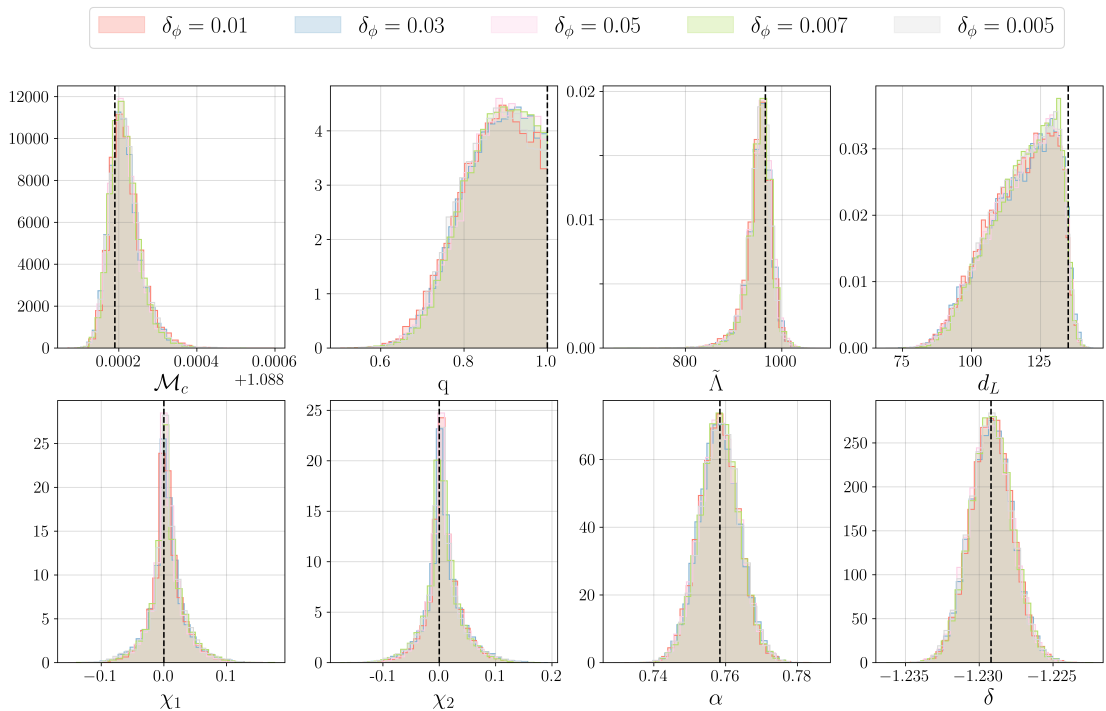}
	\caption{Comparison between the normalized posteriors for the binary parameters recovered with the \qupm{} model, for \stwo{} injections at 135Mpc with the LHVKIN network, using the relative binning technique with different precision requirements. The different colors show the posteriors for the analysis with different values of $\delta_\phi$, while the black dashed lines represent the injected values.}
	\label{fig:relbin_comp}
\end{figure*}

\section{Results for the free-parameter model}
\label{sec:free_parameter}

In the following, we show some results obtained with the postmerger model using free Lorentzian 
parameters. Performing parameter estimation analysis with the \free{} waveform requires sampling over three additional 
parameters, which implies even higher computational costs. For this reason, we could not run the same analyses with the \free{} model as we did for the \qupm{} one. 
As shown in Sec.~\ref{sec:bcs}, with high-SNR and zero-noise injections, we can recover $c_1$ 
accurately. In Fig.~\ref{fig:free_det}, we show how different detector networks can recover the 
$c_1$ parameter in the case of Gaussian noise injections, for simulated signals corresponding to 
Source3$_{\mathrm{[free-pm]}}$ at 68 Mpc. 
In the case of second generation detectors, we basically recover the prior, although with a peak between [2500,3000] Hz, where also the injected value lies. Adding NEMO to the network leads to a strong improvement, resulting in a very sharp peak for the $c_1$ posterior. The recovered value of $c_1$ with the LHVKIN network is slightly overestimated with respect to the injected value. However, this happens also for the ETCE network, where again the posterior is a sharp peak, and the injected value lies in its lower tail, outside of the $90\%$ confidence interval. In Sec.~\ref{sec:bcs}, we saw that, for the ETCE network, for the same simulated signal injected in zero noise, the value of $c_1$ is recovered very well. Therefore, we conclude that the shifts in the posterior peaks for ETCE and LHVKIN networks for the injections in Gaussian noise are most likely due to noise fluctuations, which, as reported in Sec.~\ref{sec:noise}, for this source affect also the $\tilde{\Lambda}$ measurements.
Finally, analyses of signals obtained by NR waveforms injections do not recover either of the Lorentzian 
parameters, mainly because of the complex structure of the postmerger signal in the NR waveforms, as already shown in Sec.~\ref{sec:nr}. Although the \free{} model still needs improvement for the analysis of real signals, the results in Fig.~\ref{fig:free_det} are promising, and especially show that adding NEMO to a network of second generation detectors will certainly make a difference for the study of BNS postmerger signals.

%In the case of second-generation detectors, we see a shift 
%in the peak of the posteriors for the injected value, but we find a clear improvement when adding NEMO. 
%The distribution for the ETCE network is wider but more shifted towards the correct value. We see a 
%plateau in the posterior, extending towards higher frequencies for all the networks. This feature is 
%also present for the ETCE network, for which we recover very precise posteriors in the case of zero 
%noise runs, as shown in Fig.~\ref{fig:free_bcs}. Therefore, we conclude that this is caused by noise 
%fluctuations, which, as reported in Sec.~\ref{sec:noise}, for this source affect quite heavily also the 
%$\tilde{\Lambda}$ measurements. Preliminary studies showed that changing the analysis settings, e.g., 
%fixing $c_0$ and $c_2$ to arbitrary values, helps improve the $c_1$ recovery. 
%Finally, analyses of signals obtained by NR waveforms injections do not recover either of the Lorentzian 
%parameters, because of the complex structure of the postmerger in the NR waveforms, and the noise 
%effects that we see already with simulated signals.

\section{Parameter estimation with future detectors}

\label{sec:pe}

Our discussion focused on the recovery of the $\tilde{\Lambda}$ parameter, or of the $c_1$ parameter in the case of \free{} model, because these are the quantities that encode most of the information about the EoS.
However, it is also interesting to look at the recovery of all the other binary parameters, to see how future detectors will help improving our knowledge of these systems.	
Fig.~\ref{fig:net_params} shows the comparison between the normalized posterior probability density for $\mathcal{M}_c$, $q$, $\tilde{\Lambda}$, $\chi_1$, $\chi_2$, $\alpha$ , $\delta$ and luminosity distance $d_L$, obtained using different detector networks, for \stwo{} injections at 68 Mpc and in zero noise. We find that 3G detectors will yield a strong improvement not only for what concerns $\tilde{\Lambda}$ recovery, but also in the estimation of $\mathcal{M}_c$, $q$ and $d_L$; in particular, with the ETCE network we can estimate $\mathcal{M}_c$ with a precision roughly 10 times better than the LHV one. We find only a slight improvement in the recovery of the spin magnitude values $\chi_1$ and $\chi_2$.
The best estimation of the sky location parameters $(\alpha, \delta)$ comes from the LHVKIN network, which is expected considering the larger number of detectors and their geographical distribution, as shown in Fig.~\ref{fig:det_all}.
We also note that the improvement obtained by adding NEMO to the network is roughly of a factor 1.9 and 1.6 for $\mathcal{M}_c$ and $q$ respectively, when computed in comparison with the LHVKI network, but it reaches a factor 4.4 for $\tilde{\Lambda}$ estimation. As discussed in in Sec.~\ref{sec:zeronoise}, this is achieved thanks to the postmerger contribution to the signal, which for NEMO is significant as a result of its very high sensitivity in the kilohertz band. 
Overall, future detectors will grant very precise constraints on the BNS parameters, allowing us to better understand the properties and populations of these objects. We also point out that, for computational reasons, our analyses were performed starting from a frequency $\rm f_{low} = 30\, Hz$, and hence, in reality, additional information will be available by analyzing signals starting from lower frequencies. This will lead to a large improvement especially for the 3G detectors, because, for example, the xylophone configuration of ET, with the low frequency inteferometer possibly operating at cryogenic temperatures, will ensure a good sensitivity down to $\rm f_{low} = 5\, Hz$. The additional information carried in the many inspiral cycles at low frequencies will further improve the constraints on the BNS parameters, being particularly beneficial for the spin parameters, considering that at low frequencies also spin-induced quadrupole moment effects become significant.

\section{Relative binning settings and validity}
\label{sec:relbin_app}

The relative binning method allows us to greatly reduce the computational cost of our analysis. As explained in Sec.~\ref{sec:relbin}, a fundamental requirement to employ this technique is having a reference waveform that describes the data sufficiently well. Although with real data we do not know the exact parameters of the source a priori, we can use information from low-latency analyses and quasi-universal relations to find the values to use as the fiducial parameters. Since there might still be biases in the parameters determined in such way, we checked the influence of the choice of fiducial parameters, performing some tests with different fiducial values for $\Lambda_1$,$\Lambda_2$, and we found consistency between results.

In \cite{Zackay:2018qdy}, the authors show results obtained with this method for GW170817, which, despite being a loud event, has an SNR much lower to the ones we study in this work (cf.Table~\ref{tab:snr}). The approximations used in relative binning are not expected to retain validity only in a given SNR range, but we tested the efficacy of this method applied to very loud signals by checking the consistency against results obtained with the nested sampling package \textsc{LALInference} \cite{Veitch:2014wba} of the LIGO Algorithms Library (LAL) software suite \cite{lalsuite}.

\begin{table}[]
	\setlength\extrarowheight{2pt}
	\renewcommand{\arraystretch}{1.3}
	\begin{tabular}{|c|c|c|}
		\hline
		$\delta_\phi$ & Total bins & PM bins        \\  \hline
		0.005 & 6285 & 2767 \\
		0.007 & 4489 & 1976 \\
		0.01 & 3143 & 1384  \\
		0.03 & 1049 & 462  \\
		0.05 & 630 & 277 \\ \hline
	\end{tabular}
	\caption{Number of frequency bins employed in the relative binning method for different values of $\delta_\phi$, both for the frequency range $[30,4096]$~Hz and in the postmerger region, starting at the merger frequency.}
	\label{tab:fbins}
\end{table}

Finally, when using the relative binning method, the choice of frequency bins in which the waveform is evaluated plays a crucial role. Following \cite{Zackay:2018qdy}, this choice is dictated by the requirement that the differential phase change in each bin is smaller then some threshold $\delta_\phi$. In \cite{Zackay:2018qdy}, the phase change is computed assuming a post-Newtonian (PN) description of the signal, in which the effect of the different binary parameters enter the phase with different powers of frequency. In the merger and postmerger part of the signal, the PN approximation is not valid anymore. It is not easy to find a similar way to properly describe the phase in the postmerger, without having to evaluate the waveform and incurring in computationally expensive processes that would undermine the speed-up advantage of this method. On the other hand, the phase computed with the PN approximation is then interpolated with frequency, and the frequency bins are determined by evaluating this interpolant over a grid of phases determined by the required precision $\delta_\phi$. 
Therefore, if such threshold is chosen small enough (for our analysis we set $\delta_\phi = 0.01$), we expect that the way in which the phase change is computed plays little role, and the dense frequency binning produced ensures that the bins' width is small enough to allow anyway a linear interpolation of the ratio between the generated waveform and the fiducial one, as in Eq.~\ref{eq:wfm_ratio}. 
If this was not true, we would expect that changing the threshold $\delta_\phi$, and consequently the frequency bins, over which the waveform is evaluated, would give different results also if $\delta_\phi$ was kept small. Table~\ref{tab:fbins} reports the number of frequency bins employed by the relative binning technique for different values of $\delta_\phi$, both in the whole frequency range considered for the analysis, and for the postmerger region only. Choosing small values of $\delta_\phi$ means increasing the number of bins over which we evaluate the waveform, and therefore the computational cost of the analysis; nevertheless, performing the analysis using relative binning with these settings is still much faster than running 'standard' parameter estimation analyses, which, for these kind of signals, are not computationally feasible. 
For standard parameter estimation, the waveform needs to be evaluated on a uniform grid that, with signals of the duration of roughly 200~s as the ones analysed here, includes approximately $8 \times 10^{4}$ points. Hence, considering that relative binning needs the evaluation of each sampled waveform only at the edges of the bins, this technique greatly reduces the number of required waveform evaluations.
Figure~\ref{fig:relbin_comp} shows the posteriors recovered with the \qupm{} model for the binary parameters of a \stwo injection at 135Mpc, with the LHVKIN network. We repeated the analysis multiple times, keeping the same settings but changing the frequency binning by using different values of $\delta_\phi$. We keep $\delta_\phi$ small, but look at both larger and smaller values with respect to the $\delta_\phi = 0.01$ used throughout this work. As the plot shows, we find great consistency between the results obtained with all the different values of $\delta_\phi$. Consequently, despite that the PN approximation does not hold in the postmerger phase, using it to determine the frequency bins for the relative binning method does not spoil the results, provided that the chosen $\delta_\phi$ results in small bin widths.
	
% The complex structure of the postmerger in NR waveform, together with noise fluctuations and the fact that we are sampling over three more parameters, make the Lorentzian recovery too challenging. 

%===================================================
% References

%\bibliographystyle{unsrt}
\bibliographystyle{apsrev}
\bibliography{paper}

%===================================================

%===================================================
% Acknowledgments

\end{document}